\documentclass[aps,prl,preprint,floatfix,showpacs,groupedaddress,superscriptaddress]{revtex4-1}

\usepackage{epsfig}
\usepackage{epstopdf}
\usepackage{graphicx}
\usepackage{graphics}

\usepackage{amsmath}
\usepackage{amsfonts}
\usepackage{amssymb}
\begin{document}

\title{Tunable quasiparticle trapping in Meissner and vortex states of mesoscopic superconductors}

\author{M.~Taupin}
 \affiliation{Low Temperature Laboratory, Department of Applied Physics, Aalto University School of Science, P.O. Box 13500, FI-00076 Aalto, Finland}
\author{I.~M.~Khaymovich}
\thanks{Present address: LPMMC, CNRS/Foundation Nanosciences under the aegis of Joseph Fourier University Foundation, BP 166, 38042 Grenoble, France}
 \affiliation{Low Temperature Laboratory, Department of Applied Physics, Aalto University School of Science, P.O. Box 13500, FI-00076 Aalto, Finland}
 \affiliation{Institute for Physics of Microstructures, Russian Academy of Sciences, 603950 Nizhni Novgorod, GSP-105, Russia}
\author{M.~Meschke}
 \affiliation{Low Temperature Laboratory, Department of Applied Physics, Aalto University School of Science, P.O. Box 13500, FI-00076 Aalto, Finland}
\author{A.~S.~Mel'nikov}
  \affiliation{Institute for Physics of Microstructures, Russian Academy of Sciences, 603950 Nizhni Novgorod, GSP-105, Russia}
  \affiliation{Lobachevsky State University of Nizhni Novgorod, 23 Prospekt Gagarina, 603950, Nizhni Novgorod, Russia}
\author{J.~P.~Pekola}
 \affiliation{Low Temperature Laboratory, Department of Applied Physics, Aalto University School of Science, P.O. Box 13500, FI-00076 Aalto, Finland}


%






\maketitle

{\bf
Nowadays superconductors serve in numerous applications, from high-field magnets to ultra-sensitive detectors of radiation. Mesoscopic superconducting devices, i.e. those with nanoscale dimensions, are in a special position as they are easily driven out of equilibrium under typical operating conditions. The out-of-equilibrium superconductors are characterized by non-equilibrium quasiparticles. These extra excitations can compromise the performance of mesoscopic devices by introducing, e.g., leakage currents or decreased coherence times in quantum devices.
By applying an external magnetic field, one can conveniently suppress or redistribute the population of excess quasiparticles.
In this article we present an experimental demonstration and a theoretical analysis of such effective control of quasiparticles, resulting in electron cooling both in the Meissner and vortex states of a mesoscopic superconductor.
We introduce a theoretical model of quasiparticle dynamics which is in quantitative agreement with the experimental data.
}

The presence of excess quasiparticles (QPs) is often characterized by an effective electron temperature $T$ which exceeds the temperature of the phonon bath $T_0$.
The resulting overheating
is known to be the origin of such effects as decoherence in qubit systems \cite{Martinis2009,Paik2011,Corcoles2011}, decrease of the quality factor of superconducting resonators \cite{Wang2009,Barends2011}, the excess current in single-electron turnstiles \cite{Knowles2012}, and low efficiency of electronic cooling in normal metal (N) - insulator (I) - superconductor (S) junctions \cite{Pekola2000,Rajauria2009}.
In short, overheating
is a major factor limiting the performance of S mesoscopic devices.
%
More than the overall QP number $N_{qp}$, the critical parameter is the location of these excess QPs.
For instance, for tunnel junction circuits, it is crucial to avoid the QPs in a superconductor nearby the junction, while
the extra QPs located further away are of less concern.
To suppress overheating in a superconductor one aims at lowering the generation of extra QPs in the whole superconductor using proper electro-magnetic shielding of the device, and decreasing of QP density by introducing QP traps (see e.g. \cite{Nguyen2013,Ullom1998}), by optimizing the device geometry \cite{Yamamoto2006,Knowles2012}, or by cooling using the tunnel junction to another superconductor with a larger gap \cite{Chi1979,Blamire1991,Heslinga1993}.
The second method allows one to move QPs away from critical locations and relax them.
QP traps have become an important element in designing devices for mesoscopic physics and metrology.

\begin{figure}[ht]
	\centering
		\includegraphics[width=0.48\textwidth]{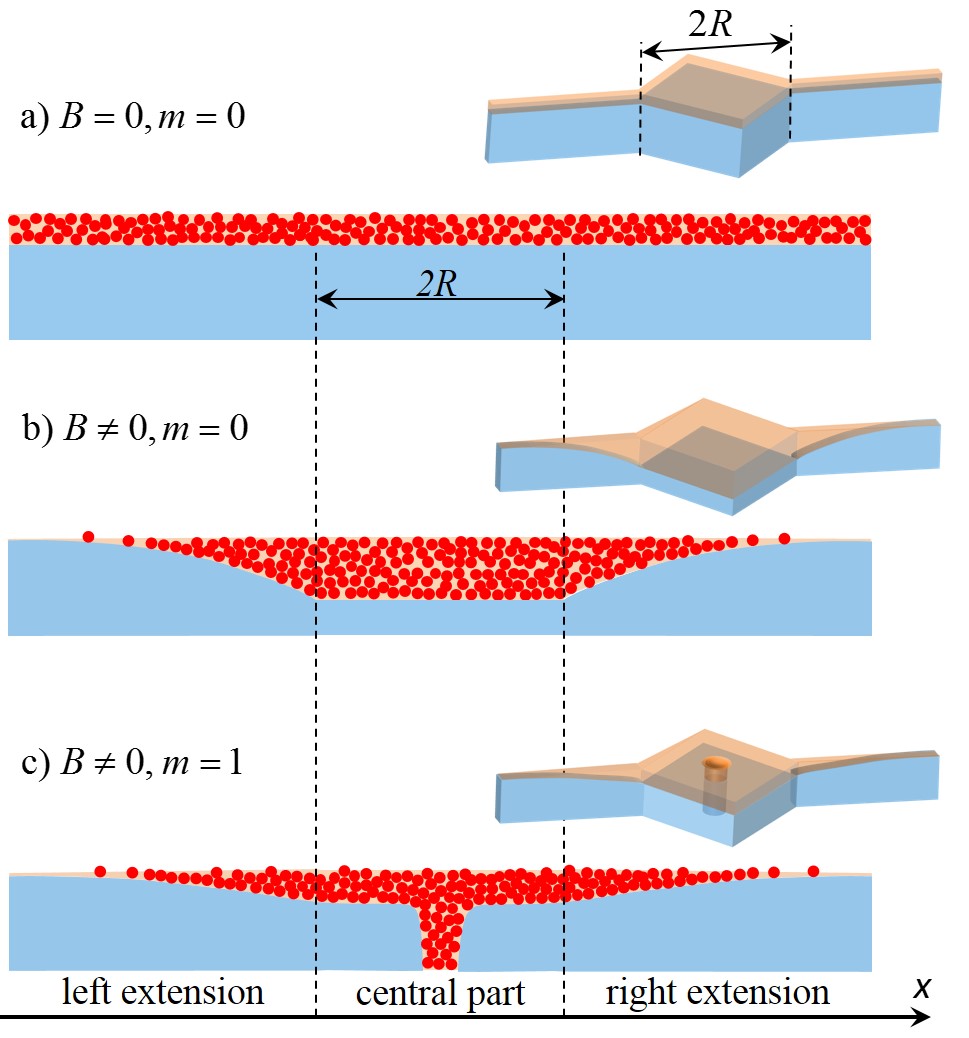}
	\caption{{\bf QP density and gap distributions in a S disc}.
The S gap $E_g(B,x)$ in a S disc with narrow extensions (Sample A) is represented by the height of the blue volume, while the QP density $n_{qp}(x)$ is shown by red circles; $m$ is the vorticity of the island, $B$ is the magnetic field acting on the sample. The wide central part of a Sample A of size $2R$ is limited by vertical dashed lines, the narrow extensions are located on the sides.
(a) Uniform zero magnetic field state; (b) Meissner state with reduced $E_g(B,x)$ in the central part at small fields in a vortex free state; (c) Single vortex state with smaller gap reduction outside the vortex core than in (b).
The 3D schematics depict the corresponding $E_g(B,x)$ (in blue) and $n_{qp}(x)$ (in orange semitransparent).
}\label{fig1_cartoon}
\end{figure}

The most common ones among different types of QP traps are normal metal sinks \cite{Goldie1990,Ullom2000,Rajauria2012}, Andreev bound states in weak links \cite{Levenson-Falk2014}, special S gap engineering \cite{Golubov1993,Golubov1994,Friedrich1997,Aumentado2004,Court2008} and non-uniform superconducting states induced by an external magnetic field \cite{Peltonen2011,Nsanzineza2014,Wang2014,Woerkom2015}.
Here we focus on the magnetic field controlled trapping, a method which has a number of advantages.
The regions with the reduced gap in this case are of the same material as the rest of the device
and therefore match perfectly the S parts without barriers or interface potentials.
Besides, magnetic field gives the possibility to make tunable traps allowing, for example, the modulation of a resonator quality factor, needed for giant pulse formation (or Q-switching) in pulse lasing (see, e.g., a book \cite{Frungel2014}).
The controllable use of such traps in various applications
mentioned above
assumes, certainly, understanding of their cooling capacities, 
 which is necessary to optimize the designing of the particular trap configurations for different mesoscopic devices. 
 
Our work aims to the solution of this ambitious and important problem  focusing on both experimental and theoretical study of individual traps which appear in the Meissner and vortex states. To build a quantitative model of these traps we choose to verify it by the experimental measurements of the characteristics of nonequilibrium
QP distributions in a mesoscopic S island (Al) in a single-electron transistor (SET) set-up with normal metal (Cu) leads.  This particular device appears to provide a very convenient way to tune both the trap pattern applying an external magnetic field to the S island
and the number of nonequilibrium QPs injected in the island in the Coulomb blockade conditions by operating it as a turnstile
of single electrons \cite{Pekola2008}.
The turnstile operation frequency $f$ of the gate voltage modulation controls QP injection rate.
This set-up allows one to probe single QP excitations in the superconducting dot by measuring the average turnstile current under pumping conditions \cite{Knowles2012} (ideally 
this current equals $ef$) and to independently control the vortex number in the superconductor \cite{Kanda2004}.
The resulting trap model has perfectly proved its validity and efficiency in this set-up which can be used in future applications.
\\

{\bf Results}

{\bf Qualitative description}.
We illustrate the key idea of QP redistribution by Fig.~\ref{fig1_cartoon}
in an S island with a large central part and two narrow extensions, called Sample A.
In the absence of magnetic field acting on the sample, $B=0$,
the QP density $n_{qp}$ is nearly uniform in the S island with constant gap $E_g(x)=\Delta_0$,
provided the heat diffusion length $L_T \gg R$ is large compared to the size of the central part $R$ (see Fig.~\ref{fig1_cartoon}(a)).
A small perpendicular magnetic field, typically few mT, which induces Meissner screening currents flowing along the superconductor edges reduces the gap $E_g(x)$ mostly in the wide central part of the island but not in the narrow extensions near the junctions \cite{Stan2004}.
Due to this non-uniform gap potential $E_g(x)$, QPs illustrated by red circles are redistributed so that their density is small at the junctions (see Fig.~\ref{fig1_cartoon}(b)).
However, the total QP number is larger than that at $B=0$ due to its exponential dependence $N_{qp}\propto e^{-E_{g,\min}/k_{\rm B} T}$ on the minimal gap $E_{g,\min}=\min_x E_g(x)$ over the island, where $k_{\rm B}$ is the Boltzmann constant.
A vortex in the island leads to further QP redistribution because it plays a role of a QP potential well containing a lot of QPs as shown in Fig.~\ref{fig1_cartoon}(c).
Despite its simplicity, the theoretical model that we present below yields a quantitative fit to the experimental data on the magnetic field and frequency dependencies of the pumping current, and thus to the QP distribution, rendering the turnstile an efficient probe of QP dynamics and relaxation.

\begin{figure}[ht]
	\centering
		\includegraphics[width=0.48\textwidth]{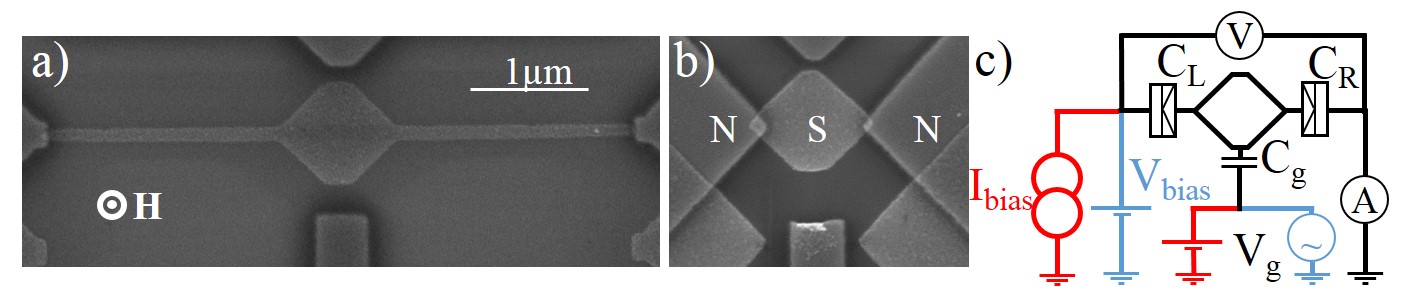}
		\includegraphics[width=0.48\textwidth]{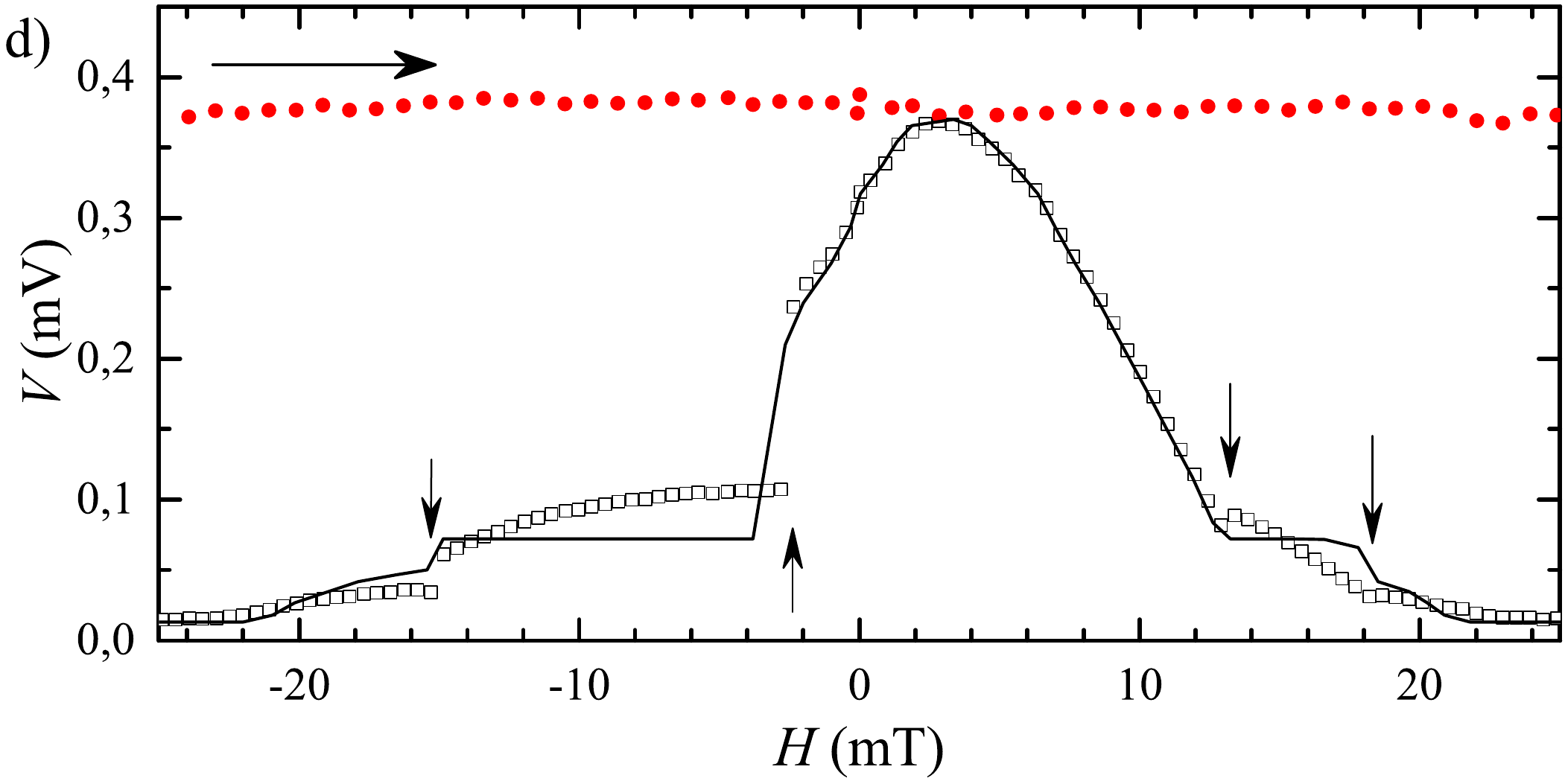}
	\caption{{\bf Layouts of the samples and results of dc measurements}.
Electron micrographs of (a) Sample A and (b) Sample B. (c) Schematic picture of the device with its electrical connections where the red and black lines correspond to the dc measurements, while blue and black ones to the pump measurements.
(d) Evolution of the voltage at a fixed bias current $I_{\rm bias}$ = 10~pA at the gate voltage $C_g V_g/e=n_g=0.5$ suppressing the Coulomb energy with the magnetic field for Sample A (red filled circles) and Sample B (black open squares).
The vertical arrows correspond to the applied field values at which the vorticity $m$ 
increases step by step by one from $-2$ to $2$ as the field is swept from $-25$~mT to 25~mT.  The horizontal arrow shows the direction of the field sweep.
The solid black line is the theoretical result for this measurement. The experimental uncertainty has been estimated as $\sim 3$~$\mu$V.}
	\label{fig2_SEM+dc}
\end{figure}


{\bf DC measurements of the S gap}.
To probe the magnetic field induced changes in the gap of a S disc, we first measure a more basic structure, which we call Sample B (see Fig.~\ref{fig2_SEM+dc}(b)). It is formed of a S disc, mimicing the central part of Sample A (Fig.~\ref{fig2_SEM+dc}(a)),
directly connected via tunnel junctions to normal leads at its edges.
Measuring electron transport through the disc while applying perpendicular magnetic field $H$ allows us to access the field dependence of the gap value $E_g(H)$ at the edge of the disc and to control the vortex state.
This way we can determine the critical fields for transitions between states with different vorticities $m$
via simple dc transport measurements (similar approach as in Ref.~\cite{Kanda2004}).
We carried out current biased dc measurements at a gate voltage that suppresses the Coulomb energy 
(for the electrical configuration, see red and black lines in Fig.~\ref{fig2_SEM+dc}(c)).
The experiments have been performed at a bath temperature of $T_{0}\sim60$~mK (well below the S gap $\Delta_0$ at $B=0$ and the Coulomb energy $E_C=e^2/(2C)$, where $C$ is the total capacitance of the island).
Note that $B$ is the actual field seen by the sample, while the applied magnetic field $H$ differs from later due to some screening by the sample holder used for shielding the sample from the environment (see Supplementary Note 1 for details).

The dc drain-source voltage $V$ measured versus the magnetic field $H$, swept from $-25$~mT to $25$~mT, is shown for Sample A (filled circles) and Sample B (open squares) in Fig.~\ref{fig2_SEM+dc}(d) at a fixed current of $I_{\rm bias}=10$~pA through the device.
In general larger voltage corresponds to larger gap $E_g(H)$ and vice versa.
The sample parameters $E_C$, $\Delta_0$, and a total normal state resistance across the two junctions $R_T$,  have been extracted from $IV$ measurements at zero magnetic field.
For the Sample B, starting from $-25$~mT, the value of the voltage is small: the island is close to its normal state.
The gap increases when decreasing the absolute value of the field till the maximum value reached at $+2.5$~mT
with two intermediate step-like anomalies at $H_{out}^{(2)}\sim -15$~mT and $H_{out}^{(1)}\sim -2$~mT, corresponding to the exit of vortices, the first one from two-vortex state to one-vortex state, and the second one from one-vortex state to a vortex free state, respectively.
Increasing $H$ further to positive values from 2.5~mT up to 25~mT leads to decrease of the gap again, with two knee-like anomalies at critical field values $H_{in}^{(1)}\sim 14$~mT and $H_{in}^{(2)}\sim 18$~mT corresponding to the entry of the first and the second vortex, respectively.
A minor distortion of the applied field (the offset in the applied field $\delta H \approx 2.5$~mT corresponding to the maximal $V(H)$ value and asymmetry of $V(H)$ in the Meissner state)
is caused by the sample-holder, and
was corrected to theoretical curves only by applying the magnetization curve $B(H)$, with $B$ the field acting on the sample, measured separately (see Supplementary Note 1 for details).
Note that the magnetic field $B$ acting on the sample itself, is zero at the maximal $V(B=0)$ and corresponds to the symmetric $V(B)=V(-B)$ in the Meissner state.
The central part of Sample A has nearly the same shape and size as Sample B; thus one can expect the critical fields of these samples to be close to each other. The anomalies are absent in Sample A, as the gap near the tunnel junctions is only weakly affected by $H$ in the presented range.
\begin{figure*}[t]
	\centering
		\includegraphics[width=0.8\textwidth]{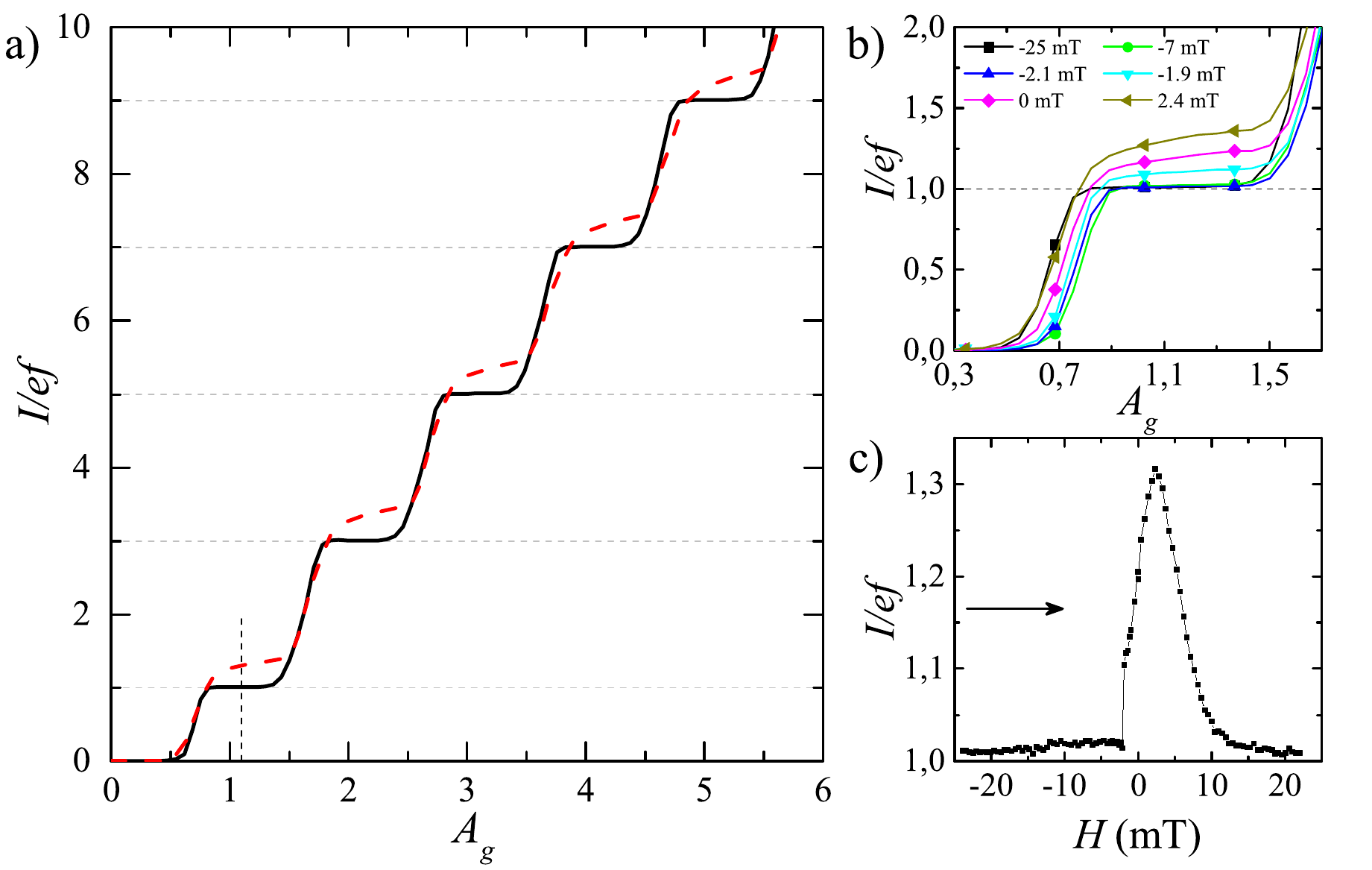}
	\caption{{\bf Turnstile current driven by an ac gate}.
(a) The pumping current $I$ normalized by $ef$ at the operating frequency $f=5$~MHz, the bias voltage $V_{\rm bias}=100$~$\mu$V, and at the gate offset $n_g^0=0.5$ versus the normalized gate amplitude $A_g$ 
at applied field $H=2.4$~mT ($B=0$) 
(red dashed curve) and at $H=-25$~mT (black solid curve). The horizontal dashed lines correspond to the expected current quantization $I =n e f$; 
(b) the zoom-up of the first plateau at several applied field values between $-25$~mT and 2.4~mT; 
(c) the evolution of the current along the vertical dashed line in the main panel versus the field $H$ varying from $-25$~mT to 25~mT; the sweep direction is shown by the horizontal arrow. In the further measurements we fixed the amplitude to the value shown by the vertical dashed line in (a). The experimental uncertainty has been estimated as $\sim 10$~fA, corresponding to $0.0125ef$ at $5$~MHz (not shown).}
	\label{fig3_I_plateaus}
\end{figure*}

{\bf Theoretical analysis of DC data}.
For the theoretical analysis of the above experimental data we simplify the standard Usadel model 
taking into account that the size of the central part $R$ of the measured samples is small compared to
the characteristic length scale of the Green's functions outside the vortex core regions (see Supplementary Note 2 for details). Such approximation leads to the Usadel equation for the normal ($\cos\theta$) and anomalous ($-i\sin\theta$) Green functions
\begin{equation}\label{Usadel_eq}
\left(iE -\Gamma\cos\theta\right)\sin\theta +\Delta\cos\theta =0 \ ,
\end{equation}
with the effective depairing parameter $\Gamma = \frac{\hbar}{2D}\langle{\bf v}_s^2\rangle$ expressed through the superfluid velocity ${\bf v}_s=D\left(\nabla\varphi - 2e{\bf A}/\hbar c\right)$ and
averaged $\langle..\rangle$ over the sample volume (over the central part of Sample A) with the excluded vortex core regions.
Here $\varphi$ is the S order parameter phase, ${\bf A}$ is the vector potential determined by the magnetic field $B$ acting on the sample, and $D$ is the diffusion coefficient. The component of ${\bf v}_s$ perpendicular to the sample boundary and to the boundaries of vortex cores should be zero.
Similarly to previous works \cite{Golubov1993,Golubov1994,Peltonen2011,Wang2014,Nsanzineza2014} the vortex cores are assumed to be normal metal cylinders of the radius $r_v$ of the order coherence length $\xi=\sqrt{\hbar D/\Delta_0}$ , i.e. $\theta=0$ inside the cores.
The sample size $2R\sim 1$~$\mu$m is also smaller than the effective screening length $\lambda_{eff}\sim\lambda^2/d_S\sim 2.6$~$\mu$m, therefore we expect uniform field distribution in the island.
Here $\lambda\sim 230$~nm \cite{Peltonen2011} is a typical bulk penetration depth and $d_S\simeq 20$~nm is the thickness of the aluminium disc.

Solution of the Usadel equation gives us the standard expression for the hard gap $E_g$ in the density of states and for the order parameter $\Delta$ as functions of $\Gamma$ (see \cite{Skalski1964,Maki1965,Fulde1965,Anthore2003} or Supplementary Note 2).
We made a fit of the field dependence of the voltage $V(B)$ at fixed currents $I_{\rm bias}$ using standard expressions for the current-voltage characteristic of a tunnel junction (see Supplementary Note 3 for details) and of the depairing parameter
\begin{equation}\label{Gamma_H_Hc}
\Gamma/\Delta_0=\alpha_1 \left(B/B_c\right)^2 - m\alpha_2 B/B_c +m^2\alpha_3 \ ,
\end{equation}
taking into account that the vector potential ${\bf A}$ in the superfluid velocity ${\bf v}_s=D\left(\nabla\varphi - 2e{\bf A}/\hbar c\right)$ is proportional to magnetic field $B$ while the S phase distribution $\varphi$ is determined by vortex sources.
Here $\alpha_i$ are numerical fitting parameters, $B_c$ denotes the field value of the first vortex entry and $m$ is the total vorticity.
The estimate $B_c\sim \Phi_0/\pi R\xi\sim 10$~mT based on $\xi\approx 100$~nm and $R\approx 0.5$~$\mu$m is rather close to the value $B(H_{in}^{(1)})\simeq14.4$~mT from our dc measurements. Here $\Phi_0=h/2e$ is the flux quantum.
More accurate estimates of $B_c$ can be done numerically, e.g., within the Ginzburg-Landau approach for a concrete sample geometry \cite{Schweigert1998, Palacios1998}.
According to \cite{Fulde1965} the parameter $\alpha_1$ determining the critical value of $\Gamma/\Delta_0$ for the first vortex to enter for the Usadel equation with homogeneous ${\bf v}_s$ in a narrow strip should be $\alpha_1^{homog}=0.237$, while the parameters $\alpha_2$ and $\alpha_3$ depend on the vortex configuration in the sample. The best fits to the experimental data are obtained with $\alpha_1=0.38$, $\alpha_2=0.438$, and $\alpha_3=0.266$, where we take $B_c=B(H_{in}^{(1)})\simeq 14.4$~mT from experimental data. 
Parameter $\alpha_1$ for a rectangular sample is expected to be a bit larger than its value $\alpha_1^{homog}$ in a narrow strip \cite{Fulde1965}.
In the fitting we assume that both jump-like and knee-like anomalies in the $V(H)$ are associated with the change of vorticity \cite{Kanda2004} and verify this applying the same parameters to $V(H)$ with different values of $I_{\rm bias}$ (see Supplementary Note 3).
The S gap in the narrow extensions of Sample A shown in Fig.~\ref{fig2_SEM+dc}(d) is close to its zero-field value $\Delta_0$ up to $\sim30$~mT with few \% accuracy as the depairing parameter in this case $\Gamma/\Delta_0 = (\pi \xi w B/\Phi_0)^2/6$ is small \cite{Anthore2003}. Here $w\simeq 130$~nm is the width of the extensions.

{\bf Pumping measurements}.
The pumping measurements are done on Sample A which has the highly non-uniform distribution of the gap (Fig.~\ref{fig1_cartoon}) under magnetic field.
To probe the magnetic field dependence of non-equilibrium QP states, we measure the current $I$ in turnstile mode averaged over the period of the drive $\tau_0=1/f$ \cite{Pekola2008}.
We apply a fixed bias voltage $V_{\rm bias}=100$~$\mu$V and sinusoidal gate voltage $C_g V_g/e = n_g(t) = n_g^0+A_g \sin(2\pi f t)$ through the capacitor $C_g$ with variable amplitude $A_g$.

\begin{figure}[t]
	\centering
		\includegraphics[width=0.48\textwidth]{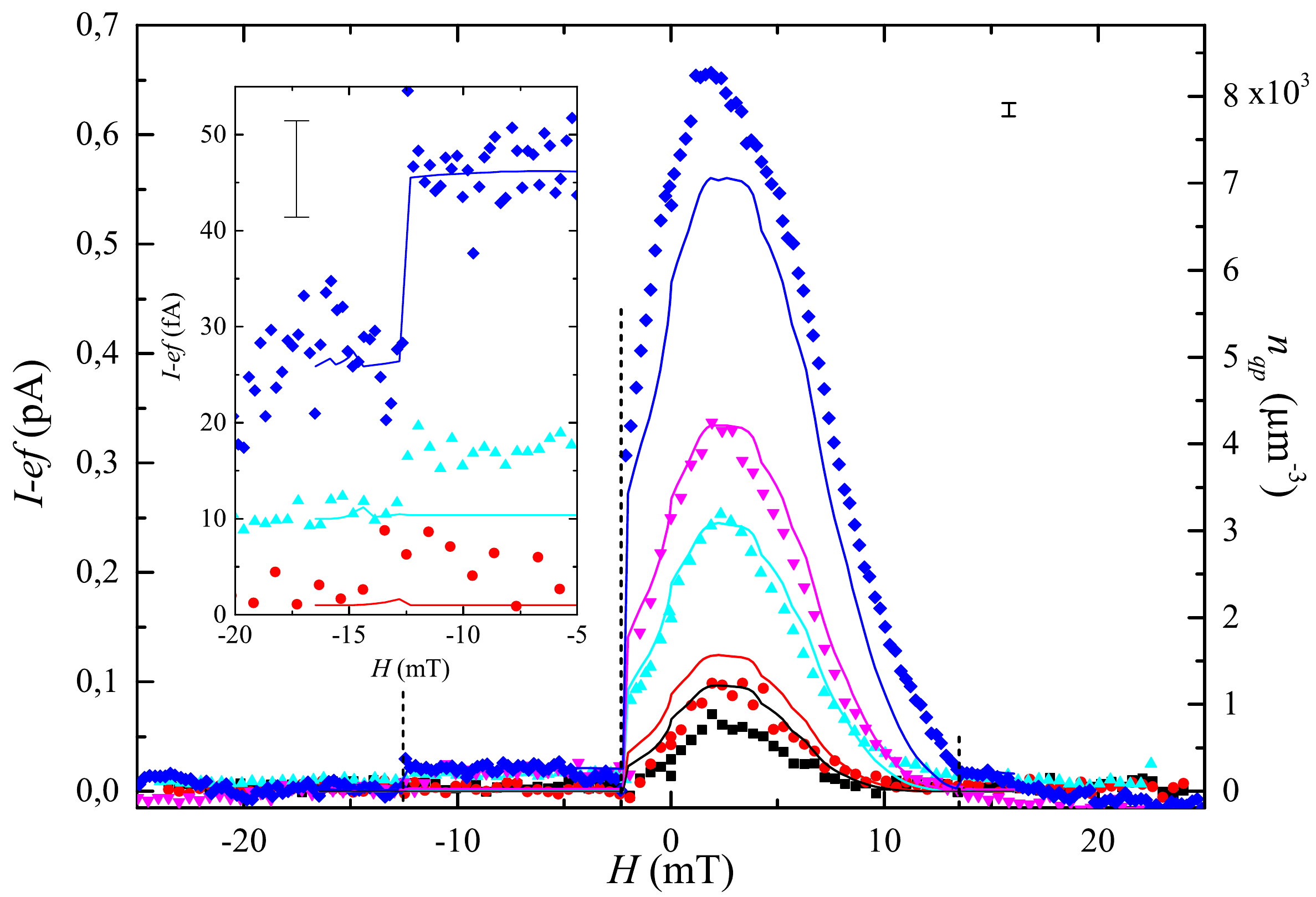}
	\caption{{\bf Excess pumping current vs magnetic field}.
The excess current $I-ef$ at the first plateau for the driving frequencies $0.5$~MHz (black), $1$~MHz (red), $5$~MHz (cyan), $10$~MHz (violet), and $30$~MHz (blue) with the maximum electronic temperature at $B(H)=0$ of 274, 286, 320, 337, and 369\,mK respectively.
Applied magnetic field $H$ is swept from $-25$~mT to 25~mT (the sweep direction is shown by a horizontal arrow).
The vertical dashed lines show the expected values corresponding to the entrance of an extra vortex into the island (for $H>0$) and to the removal of one vortex ($H<0$). (inset) A close-up of the current for a field close to the expulsion of the second vortex.
For better visibility, the data sets for different frequencies have been shifted vertically.
The measurements were performed at the bias voltage $V_{\rm bias}=100$~$\mu$V, the gate offset $n_g^0=0.5$. 
In all panels symbols (solid lines) correspond to the experimental data (theoretical model) with the estimated maximal QP temperature shown in numbers.
The scale on the right side shows the QP density $n_{qp}$ near the junction.
The experimental uncertainty is $\sim 10$~fA, shown as an error bar on both panels.
}
	\label{fig4_dI_vs_B}
\end{figure}

The turnstile current is expected to assume values equal to integer multiples of $ef$ in the absence of non-equilibrium effects and unwanted tunneling events.
The measurements are carried out around the gate offset point $n_g^{0}=0.5$, to maximize the expected plateau width, for several frequencies $f$.
Overheating of the S island, in particular at $H=2.4$~mT corresponding to $B(H)=0$, leads to positive deviations of this current from $ I =nef$ ($n$ is an integer) by tens percents at the expected plateau positions which corresponds to thousands of QPs per $\mu$m$^3$ near the junction (see Fig.~\ref{fig3_I_plateaus}(a) at $f=5$~MHz).
The magnetic field improves QP trapping: the deviation from $ef$ (and the corresponding QP density) at large enough magnetic fields decreases by an order of magnitude in the whole frequency range from 0.5 to 200~MHz (see Fig.~\ref{fig4_dI_vs_B} and Supplementary Note~3) and approaches a few-percent level related to an amplifier noise, even for large gate amplitudes when pumping up to $n=11$ electrons per cycle.
The zoom up of the first plateau shown in Fig.~\ref{fig3_I_plateaus}(b) demonstrates the magnetic field dependence of $I$.
To separate the Meissner current from the vortex contribution, we present in Fig.~\ref{fig3_I_plateaus}(c) pumping current versus the field at a fixed gate amplitude value indicated by the vertical dashed line in the main panel.
The excess current $\delta I=I-ef$ increases when the field is swept from large negative values to low values with jumps at $H_{out}^{(1,2)}$ (see vertical dashed lines in Fig.~\ref{fig4_dI_vs_B}).
The following field increase to positive values leads to decreasing excess current without visible anomalies. This is related to the difference in the $k$th vortex entry (exit) fields $H_{in(out)}^{(k)}$.
Indeed, 
for $k=1$ at these fields, we have $E_g(H_{in}^{(1)})< E_g(H_{out}^{(1)})$
which leads to the efficient redistribution of QP density even without any vortex (see Fig.~\ref{fig1_cartoon}(b)).
Despite the absence of anomalies at the vortex entries, it is possible to estimate the value $H_{in}^{(1)}$ by varying the value of the initial field: the discontinuous anomaly at $H_{out}^{(1)}$ is only visible for a field amplitude in a sweep exceeding $H_{in}^{(1)}=13.5$~mT which is close to the value found by dc measurements in Sample B.
At even higher values of the field, $|H|\gtrsim 30$~mT, the current quantization is lost again due to the eventual suppression of the S gap near the junctions as well (see Supplementary Figs.~4).

{\bf Theoretical analysis of pumping data}.
To model theoretically the excess current as a function of the field $B$ and frequency $f$ we calculate the electronic temperature $T$ using a heat balance equation
\begin{gather}\label{heat_balance_eq}
\dot Q_{eph}(T)\simeq I V_{\rm bias} = [e f+\delta I(T)] V_{\rm bias} \ .
\end{gather}
We keep in mind that $T$ is nearly uniform and constant in time provided the heat diffusion length $L_T$ is large compared to the size of the island $R$, and the heat relaxation time $\tau_{eph}$, determined by electron-phonon coupling, is much larger than the operating period $\tau_0=1/f$, i.e., $\tau_{eph}\gg \tau_0$ allowing us to average the heat diffusion equation across the sample volume and over the operating period.
We assume further that most of the Joule dissipation occurs inside the S island and take into account the excess pumping current $I$ averaged over the period from the first plateau on the right hand side (see Supplementary Note 5 for details).
The value of $\dot Q_{eph}=\dot Q_{eph}^{(nv)}+\dot Q_{eph}^{(v)}$ describes the heat flow rate from the electronic subsystem to phonons for a non-zero depairing parameter $\Gamma$.
We calculate its value $\dot Q_{eph}^{(nv)}(T)$ outside the vortex core regions for any $\Gamma$ combining the procedure described, e.g., in Refs.~\cite{Kopnin2001,Timofeev2009} and the solution of Eq.~\eqref{Usadel_eq}. For the experimental parameters $k_{\rm B} T_0, \Gamma^{2/3} E_g^{1/3}\ll k_{\rm B} T\ll E_g$ in low-temperature limit the electron-phonon heat flux
\begin{gather}\label{Q_eph}
\dot Q_{eph}^{(nv)}(T) \simeq \frac{\Sigma (\mathcal{V}-\mathcal{V}_v)}{\zeta(5)}\left\{ \frac{64}{63} T^5 e^{-\frac{E_g}{k_{\rm B} T}}
+ \frac{2\pi E_g^4}{3k_{\rm B}^4} T e^{-\frac{2E_g}{k_{\rm B} T}}\right\}
\end{gather}
decomposes into recombination $\propto e^{-\frac{2E_g}{k_{\rm B} T}}$ and scattering terms $\propto e^{-\frac{E_g}{k_{\rm B} T}}$ (see, e.g., \cite{Maisi2013}).
Here $\Sigma$ is the electron-phonon coupling constant, and $\mathcal{V}$ ($\mathcal{V}_v$) is the volume of the island (the vortex core regions).
In the vortex cores in the same limit of negligible phonon temperature $T_0\ll T$ the electron-phonon heat flow is modelled by the standard normal metal expression $\dot Q_{eph}^{(v)}=\Sigma \mathcal{V}_v T^5$  with the volume of $m$ vortex cores assumed to be $\mathcal{V}_v = m r_v^2 d_S$. The recombination term in Eq.~\eqref{Q_eph} becomes dominant at $k_{\rm B} T\gtrsim 0.1 E_g$. Beyond the low temperature limit we use a numerically calculated expression for $\dot Q_{eph}^{(nv)}(T)$ instead of \eqref{Q_eph} (see Supplementary Note 5 for calculation details).

Eventually we obtain the magnetic field and frequency dependence of the measured excess current $\delta I(T)$ as
\begin{gather}\label{dI(T)}
\delta I(T) = C\frac{\sqrt{2\pi \Delta_0 k_{\rm B} T}}{e R_T}\exp[-\Delta_0/k_{\rm B} T] \ .
\end{gather}
Note that the QP density near the junction is proportional to the excess current $n_{qp}= D(E_F) e R_T \delta I(T)/C$ and can be extracted from $\delta I(T)$ using the normal state density of states in the superconductor $D(E_F)$ (see the scale on the right side of Fig.~\ref{fig4_dI_vs_B} showing the QP density $n_{qp}$).
Here $C\sim 1$ is a numerical coefficient determined by the wave-form and the amplitude $A_g$ of the gate drive, in particular the duration for one junction to be open for tunneling in each cycle.
A detailed derivation is given in Supplementary Note 6.
Note that in Eq.~\eqref{dI(T)} we neglected contributions of higher order processes in $R_T^{-1}$ like Andreev tunneling due to the small transparency of the junctions (see experimental results in \cite{Aref2011} and estimates in Supplementary Note 6).
By solving \eqref{heat_balance_eq} with the substituted expressions (\ref{Gamma_H_Hc}, \ref{dI(T)})
we find the solution for $T$ and $\delta I(T)$ (solid lines in Fig.~\ref{fig4_dI_vs_B}).
We used the constant $C=1$ for a fixed drive amplitude $A_g$.
The main uncertainty in the fitting procedure originates from the parameter $\Sigma \mathcal V$.
The volume of the S sample can be estimated based on the electron micrograph (see Fig.~\ref{fig2_SEM+dc}(b)) as $\mathcal V\simeq 3\cdot 10^{-20}$~m$^3$, but usually this value is overestimated due to additional uncontrolled oxidation of Al. 
On the other hand the typical range of the measured values of the electron-phonon relaxation constant $\Sigma$ in the bulk aluminium \cite{Kautz1993,
Giazotto2006,Maisi2013} is within $2\cdot 10^8$ to $5\cdot 10^8$~W K$^{-5}$ m$^{-3}$.
Our fitting gives results agreeing reasonably well with the experimental data within the range of $\Sigma \mathcal V$ from $4\cdot 10^{-12}$ to $9\cdot 10^{-12}$~W K$^{-5}$.
In Fig.~\ref{fig4_dI_vs_B} we present a fit for a certain middle value $\Sigma \mathcal V=6\cdot 10^{-12}$~W K$^{-5}$ which is in the best agreement 
with the experiment at moderate frequencies. Assuming $\mathcal V\simeq 3\cdot 10^{-20}$~m$^3$ we get $\Sigma=2\cdot 10^8$~W K$^{-5}$ m$^{-3}$ which is towards the low end due to the overestimated $\mathcal V$ but within the range given above.
 We have extracted the optimal value of the vortex core radius within the range $r_v=2.5-2.7\xi$ both from the dc measurements (see Supplementary Note 3) and from the pumping data, which is in perfect agreement with the previous theoretical results \cite{Golubov1993,Golubov1994}.
\\

{\bf Discussion}

According to the theoretical model, Eq.~\eqref{dI(T)}, the maximal electronic temperature at 
$A_g=1.1$ and $f=30$~MHz is $T\simeq370$~mK. It corresponds to a number of non-equilibrium QPs $N_{qp}=n_{qp} \mathcal{V}\simeq 250$ in the uniform state (see Fig.~\ref{fig1_cartoon}(a)).
In the field increasing from $B(H)=0$ the Meissner supercurrents sufficiently improve the electron-phonon relaxation by reducing the gap $E_g(\Gamma)$ in the central part of the island even before the first vortex enters the island. This leads to the at least $10-20$ times reduction of the QP density near the junction when the excess current approaches the amplifier noise level.
The vortex contribution is clearly seen in the decreasing field regime due to the hysteresis caused by vortices.
Indeed, the vortices that entered the island at a certain value of the field stay there till smaller fields (where the effect of Meissner current is smaller) and improve the relaxation of hot QPs most effectively.
Such hysteresis allows us to see the vortex contribution alone (see the larger step in Fig.~\ref{fig4_dI_vs_B} at $H\sim-2$~mT) and the improvement of relaxation in the two-vortex state with respect to the one-vortex state (the smaller step at $H\sim-13$~mT).
We estimate the recombination rate in the vortex state $\Gamma_{rec}\simeq f/N_{qp,vort}$ as the injection QP rate $f$ divided by the QP number $N_{qp,vort}\simeq 2 D(E_F) \mathcal{V}_v k_{\rm B} T \ln 2$ in the vortex core volume $\mathcal{V}_v$ (see Supplementary Note 5 for details).
At $f=30$~MHz it gives $N_{qp,vort}\sim 100$, $\Gamma_{rec}\simeq 0.3$~MHz of the recombination rate, i.e., $20$ times higher than $\Gamma_{rec}^0=16$~kHz estimated in \cite{Maisi2013} at $B=0$.

In conclusion, we demonstrate effective control of the number of excess quasiparticles and their spatial distribution in a mesoscopic superconducting disc by applying a small magnetic field on it.
We find that both the Meissner supercurrents and vortices entering the disc one by one each give important observable contributions to the trapping of non-equilibrium quasiparticles.
We demonstrate that a single-vortex contribution is sufficient to keep the superconducting disc near equilibrium up to $30$~MHz injection frequency
with $n_{qp}\simeq 400$~$\mu$m$^{-3}$ quasiparticle density near the junction and recombination rate of order of $\Gamma_{rec}\simeq 0.3$~MHz. Our dc and pumping measurements confirm the assumption \cite{Golubov1993,Golubov1994} that a vortex can be considered as a normal metal cylinder with the effective radius $r_v=2.5-2.7\xi$ both in charge and heat transport problems.
Our theoretical analysis of the quasiparticle trapping has proven its validity and efficiency in the set-up being in quantitative agreement with the experimental data.

\subsection*{Methods}
{\bf Device fabrication}.
The hybrid devices with aluminium as the superconductor,
copper as the normal metal, and aluminium oxide as the tunnel
barrier in between, have been fabricated by standard electron-beam lithography and two-angle shadow evaporation technique.
The aluminium island is $d_S = 20$~nm thick and it is oxidized with $O_2$ for $2$~min at $2$~mbar.
The copper leads, $25$~nm thick, are placed on the oxidized Al forming tunnel junctions.

{\bf Sample geometries and parameters}.
Two different island geometries have been employed in the measurements: Sample B has a nearly square-shaped island, as shown Fig.~\ref{fig2_SEM+dc}(b), and Sample A with the same central part as geometry B has two additional long narrow aluminium extensions from each side toward the junctions (Fig.~\ref{fig2_SEM+dc}(a)).
The diagonals of the island are $2R\sim$1~$\mu$m both in A and B, and the narrow extensions of the island in A are $2$~$\mu$m long and $w\sim 0.13$~$\mu$m wide.
The sum of the tunnel resistances of the two junctions is $R_T\simeq 577$~k$\Omega$ for Sample A and $R_T\simeq 714$~k$\Omega$ for Sample B.
We measured the $IV$ characteristics of SETs at various values of the DC gate voltage at the base temperature to determine the zero field S gap value $\Delta_0\simeq 190$~($207$)~$\mu$eV
and the charging energy $E_C\simeq 173$~($133$)~$\mu$eV for Sample A (B).

{\bf Reproducibility and noise}.
All the results presented here are reproducible between different runs and between samples of similar geometry, in particular, as concerns the values of the critical fields of vortex entry (exit).
The results depend only on whether the absolute field value increases or decreases,
provided by the hysteresis in vortex entry/exit events, but they do not depend on the sign of the field as such. 
The samples are cooled down through the superconducting transition with a zero-field cooled magnet. The uncertainties of current and voltage are estimated to be 10\,fA and 3\,$\mu$V respectively. They are taken as the noise from the amplifiers.


%

\subsection*{Acknowledgements}
We want to thank V. Maisi for valuable discussions, and T. Faivre for technical help and useful comments.
This work has been supported in part by Academy of Finland 
(Project Nos. 284594, 
272218), 
by the European Union Seventh Framework Programme INFERNOS (FP7/2007-2013) under Grant Agreement No. 308850, by Microsoft Project Q, by the EMRP (project no. SIB01-REG2), by the Russian Foundation for Basic Research, and the grant of the Russian Ministry of Science and Education No. 02.B.49.21.0003.
We acknowledge the availability of the facilities and technical support by Otaniemi research infrastructure for Micro and Nanotechnologies (OtaNano).

\subsection*{Author Contributions}
M.~T., M.~M., and J.~P.~P. conceived and designed the experiments;
M.~T. performed the experiments;
M.~T., I.~M.~K., and A.~S.~M. analyzed the data.
M.~T., I.~M.~K., M.~M., A.~S.~M., and J.~P.~P. contributed with materials/analysis tools;
M.~T., I.~M.~K., M.~M., A.~S.~M., and J.~P.~P. wrote the paper.

\subsection*{Additional Information}
Supplementary notes are available in the online version of the paper.
Reprints and permissions information is available at www.nature.com/reprints.
Correspondence and requests for materials should be addressed to M.~T.~(email: mathieu.taupin@aalto.fi).

\subsection*{Competing Financial Interests}
 The authors declare that they have no
competing financial interests.

\newpage
\part{Supplementary Materials}

\section*{Supplementary Note 1. Hysteresis in the measurements under field}\label{SM_sec1:B(H)}

As pointed out in the main text, a remanent field of $\delta H \approx 2.5$~mT and the asymmetry in the Meissner state are present at the sample location at zero applied field $H=0$ due to the presence of superconducting parts in the sample-holder. The dc measurement presented Fig. 2(d) of the main text for Sample B has been reproduced (on the same sample) in a sample-holder that does not have the superconducting shield for which $B=H$. These measurements are shown in Supplementary Figure~\ref{fig1SM_dc_V_vs_B} for several bias current values $I_{\rm bias}=1$ (blue triangles), $10$ (red circles), and $100$~pA (black squares), together with the theoretical model for the experimental data (solid lines of corresponding colors). 
The dc voltage $V$ vs field $B$ in this measurement is symmetric with respect to the zero applied field value (except for the vortex hysteresis intrinsic for the sample). The theoretical model presented in Supplementary Note 3 reproduces perfectly the experimental points.

\begin{figure}[ht!]
	\centering
		\includegraphics[width=0.5\textwidth]{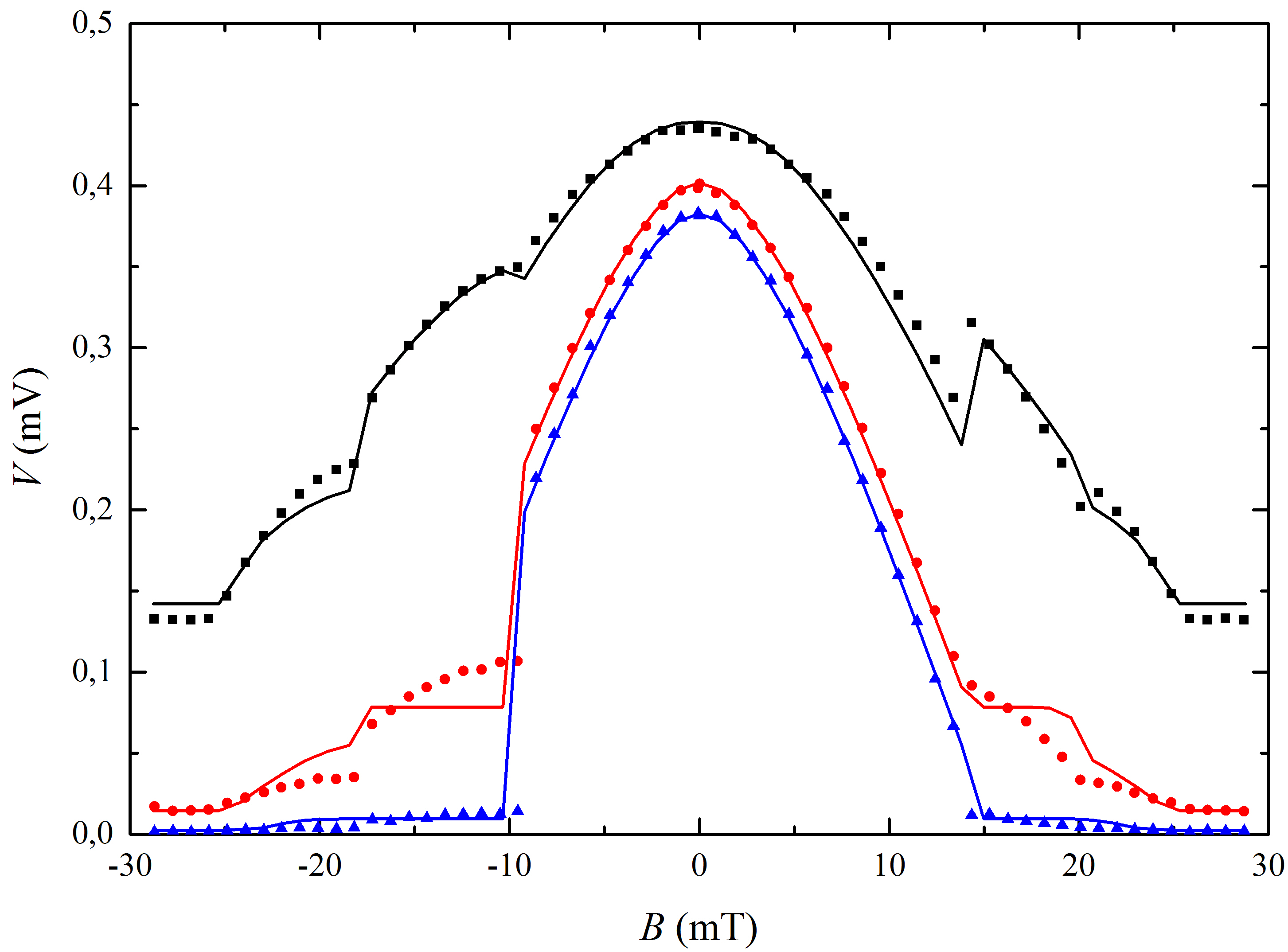}
	\caption{{\bf DC measurements without field distortion.} Evolution of the voltage at a fixed bias currents $I_{\rm bias}=1$ (green triangles), $10$ (red circles), and $100$~pA (black squares) at the gate voltage $C_g V_g/e=n_g=0.5$ suppressing the Coulomb energy with the magnetic field for Sample B in a sample-holder that does not have the superconducting shield together with the theoretical model (solid lines of corresponding colors). The field $B$ is swept from $-30$~mT to 30~mT.}
	\label{fig1SM_dc_V_vs_B}
\end{figure}

In order to fit the theoretical model to our measurements versus field $H$ performed in the sample-holder with the superconducting shield (i.e. with deformation of the field profile), a correction has to be done, as the field $H$ applied through the coil differs from the effective ``acting'' field $B$ seen by the sample.
The field profile has been measured in the shielded sample-holder with a Hall sensor at 4.2~K and at 0.2~K.
The first measurement has been done in the normal state when there is no magnetic shielding as a reference point (not shown).
The second measurement of the effective field $B$ versus the applied coil magnetic field $H$ swept from $-30$~mT to 30~mT (at 0.2~K) is shown in Supplementary Figure~\ref{fig2SM_Hcorr} as black dashed lines.
The arrows point out the direction of the sweep.
For $|H|>20$~mT, the sample-holder is fully normal and the effective field equals the applied one.
For $|H|<20$~mT, a nonlinear superconducting response from the sample holder is present leading to hysteresis.
The red line in Supplementary Figure~\ref{fig2SM_Hcorr} is the correction found by comparing directly the dc measurements in both sample-holders ($V(H)$ in Fig.~2(d) of the main text and $V(B)$ in Supplementary Figure~\ref{fig1SM_dc_V_vs_B}).
The two methods yield very similar results.
Eventually theoretical curves given as functions of $B$ by the model are presented as functions of $H$ using red curve $B(H)$ in Supplementary Figure~\ref{fig2SM_Hcorr}.

\begin{figure}[ht!]
	\centering
		\includegraphics[width=0.5\textwidth]{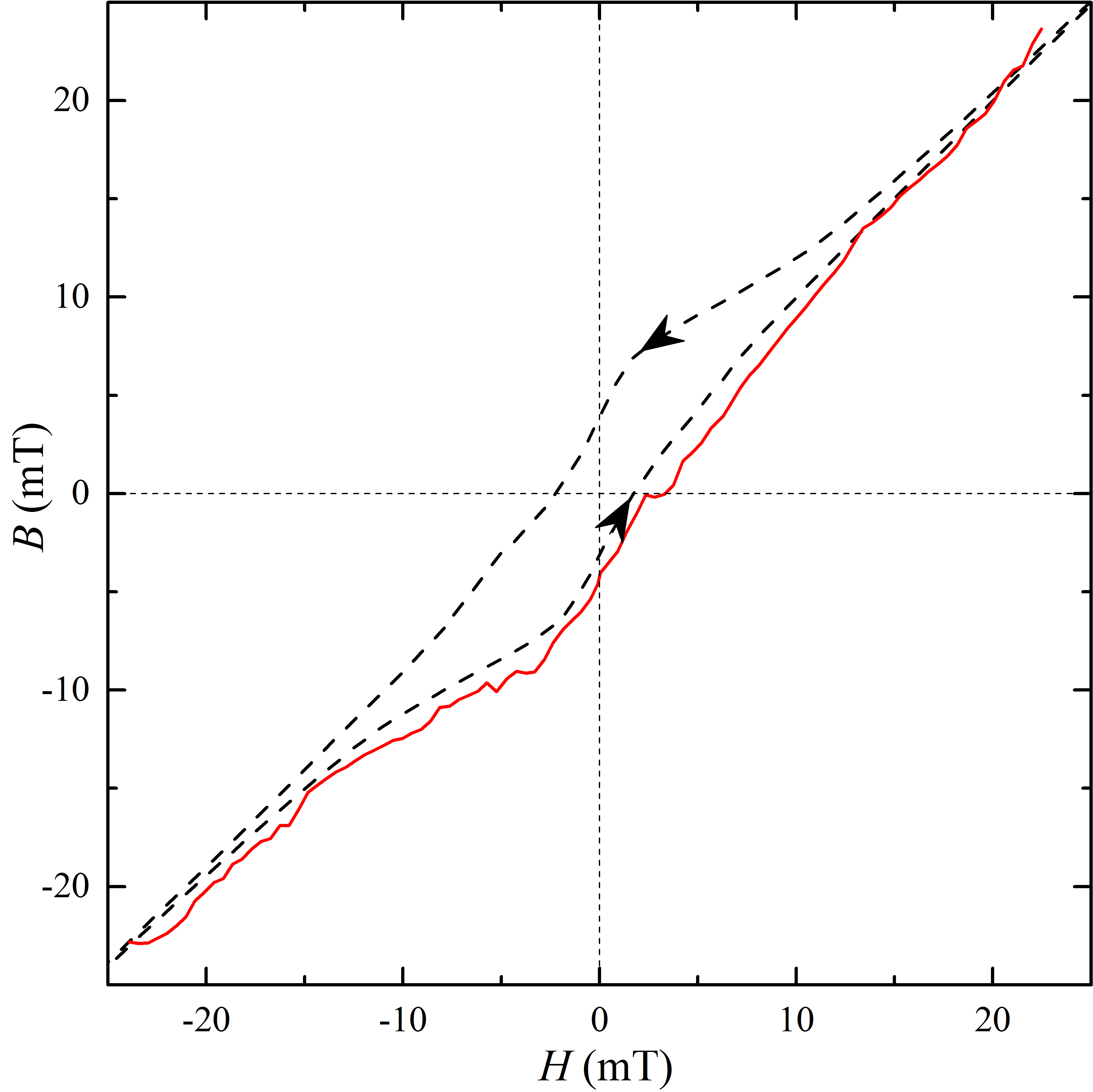}
	\caption{{\bf Magnetization curves.} Field correction found by using a Hall sensor (black dashed lines) and by comparing the dc measurements of $V(H)$ in both sample holders (red solid line). The largest deviation from linearity lies at low field. The two methods yield a very similar correction.}
	\label{fig2SM_Hcorr}
\end{figure}

\section*{Supplementary Note 2. Homogeneous approximation in a mesoscopic sample}\label{SM_sec2:homog_Usadel}

The samples studied experimentally are in the dirty regime, namely $\ell\ll \xi$, where $\xi=\sqrt{\hbar D/\Delta_0}$ is the coherence length, $D$ is the diffusion coefficient, $\ell$ is the elastic mean free path, and $\Delta_0$ is the superconducting gap.
The quasiparticle (QP) spectral characteristics in this case can be found from the Usadel equations (see, e.g., \cite{Anthore2003})
\begin{equation}\label{Usadel_eq}
\frac{\hbar D}{2} \nabla^2\theta({\bf r}) +\left(iE -\frac{\hbar}{2D}{\bf v}_s^2\cos\theta({\bf r})\right)\sin\theta({\bf r}) +\Delta\cos\theta({\bf r}) =0 \ ,
\quad {\rm div}\left( \sin^2\theta({\bf r}) {\bf v}_s \right) =0\ ,
\end{equation}
where $g^R=-g^{A*}=\cos\theta({\bf r})$ and $f^{R}=-f^{A*}=-i\sin\theta({\bf r})$ are the normal and anomalous Green's functions (superscripts 'R' and 'A' stand for 'retarded' and 'advanced').

Considering the experimental situation of a mesoscopic superconducting sample with the characteristic size $R=0.5$~$\mu$m and the coherence length $\xi\sim 100-200$~nm (depending on the diffusion coefficient), we have to verify if we can neglect the gradient terms in Sup. Eq.~\eqref{Usadel_eq}.
For subgap energies the characteristic length scale of the function $\theta({\bf r})$ can be estimated as follows: $\xi/\sqrt{1-E^2/\Delta^2}$.
It is natural to assume the $\theta({\bf r})$ function inhomogeneity to be small provided $\xi/\sqrt{1-E^2/\Delta^2}>R$.
This condition gives us the energy interval $1-E^2/\Delta^2<\xi^2/R^2\sim 0.1$, sufficient for the calculations of the electron-phonon heat flow $\dot Q_{eph}$ and the thermal excitation leakage current $\delta I$ for temperatures much lower than the superconducting gap. Indeed, the main contribution to $\dot Q_{eph}$ and to $\delta I$ is given by $|E/\Delta-1|\sim k_{\rm B} T/\Delta\simeq 0.03 < \xi^2/R^2\simeq 0.1$.

Certainly the above assumption is strictly valid only for the Meissner state: inside the vortex core the gap and the anomalous Green function turn to zero at the scale of  the effective core radius  $r_v$ which is of order of the coherence length $\xi$ \cite{Golubov1993}.

To avoid numerical solution of the Usadel equation we adopt in the main text the following approximate procedure.
In the presence of vortices we assume that both the order parameter and $\theta({\bf r})$ function vanish inside the vortex cores while outside the core regions
we assume the $\theta({\bf r})$ function to vary slowly and introduce, thus, its average $\theta$ over the region outside the vortex cores (omitting the spatial dependence in the notation).
The deviations from the averaged order parameter $\Delta$ beyond the cores also become small in this limit.
Integrating now the above Usadel equation over the region outside the vortex cores we obtain Eq.~(1) from the main text with
the effective depairing parameter expressed through the superfluid velocity ${\bf v}_s$ as
\begin{equation}
\Gamma = \frac{\hbar}{2D}\langle{\bf v}_s^2\rangle =\frac{\hbar D}{2}\langle\left(\nabla\varphi - 2e{\bf A}/\hbar c\right)^2\rangle  \ .
\end{equation}
Here the brackets $\langle..\rangle$ denote an average over the sample volume (over the central part of the sample A) with the excluded vortex core regions,
$\varphi$ is the superconducting order parameter phase and ${\bf A}$ is the vector potential determined by the external magnetic field $B$ applied to the sample.
The second Usadel equation in our approximation reduces to ${\rm div} {\bf v}_s =0$ and leads to
the vanishing components of ${\bf v}_s$ perpendicular to the sample boundary and to the boundaries of vortex cores.
Here and further on we neglect the changes in the magnetic field $B$ due to the screening currents flowing in the sample due to the smallness of the characteristic sample size $R$ as compared to
the effective screening length $\lambda_{eff}=\lambda^2/d_S$. For our samples $\lambda\simeq 230$~nm \cite{Peltonen2011} and $d_S=20$~nm, therefore $\lambda_{eff}\simeq 2.6$~$\mu$m.

Solution of the averaged Usadel equation gives us the expression for the hard gap $E_g$ in the density of states and for the order parameter $\Delta$ as functions of $\Gamma$ as \cite{Skalski1964,Maki1965,Fulde1965,Anthore2003}
\begin{gather}
E_g = \Delta (1-\gamma^{2/3})^{3/2} \ , \; \Delta = \Delta_0 e^{-\pi \gamma/4} \ , \; \gamma = \Gamma/\Delta \ .
\end{gather}
In the main text we focus on the case $\gamma<1$ ($\Gamma<\Delta_0 e^{-\pi/4}$), implying that the gap $E_g>0$ is non-zero.

\section*{Supplementary Note 3. DC fitting}\label{SM_sec3:dc_fit_V(H)}
Using the solution of the averaged Usadel equation, Eq.~(1) from the main text, one can fit the $IV$ characteristics shown in Supplementary Figure~\ref{fig1SM_dc_V_vs_B}.
Indeed, we consider a hybrid single electron transistor (SET), namely, a mesoscopic superconducting island tunnel coupled to the normal metal leads (NISIN).
We apply a fixed bias current $I_{\rm bias}$ through the normal leads and the constant gate voltage $n_g = C_g V_g/e = 0.5$ to the gate electrode coupled to the island through the capacitor $C_g$ (see black and red lines in Fig.~2(c) of the main text) and measure the difference $V$ of voltages $V_{L,R}=\pm V/2$ applied to the leads as a function of the magnetic field $B$ seen by the sample.

In stationary state the current $I_{\rm bias}$ flowing from one lead to another is equal in any cross section and it can be calculated in any of two junctions (for example, in the left one)
\begin{gather}\label{I_pk}
I_{\rm bias}=-e \sum_k p_k \left[\Gamma_{k\to k+1}^L(V)-\Gamma_{k\to k-1}^L(V)\right] \ .
\end{gather}
as a sum over the island charge state $k$ of the sequential tunneling rates $\Gamma_{k\to k+1}^L$ ($\Gamma_{k\to k-1}^L$) to (from) the island through the left junction.
This sum is weighted with the probability $p_k$ of system being in this charge state, which is calculated using the standard rate equation for the balance of the probability fluxes \cite{Averin1986,Fulton1987,Likharev1987}
\begin{gather}\label{master_eq_k}
\frac{d p_k}{dt} = \sum\left[\Gamma_{k\pm1\to k} p_{k\pm1}-\Gamma_{k\to k\pm1} p_k\right] \ , \quad \sum p_k = 1 \ ,
\end{gather}
in the stationary case ${d p_k}/{dt}=0$ with the tunneling rates
$\Gamma_{k\to k\pm 1} = \sum_{i=L,R}\Gamma_{k\to k\pm 1}^i$ and $\Gamma_{k\to k\pm 1}^i=\Gamma[U_{k,i}^{\pm}]$ given by
\begin{gather}\label{Gamma[U]}
\Gamma[U]=\frac{2}{e^2 R_T}\int n_S(E)f_T(E)\left[1-f_{T_0}(E+U)\right]dE \ .
\end{gather}
Here $U_{k,i}^{\pm}=\mp 2E_C(k-n_g\pm1/2)\mp eV_i$ are the energies gained by the electron tunneling to/from the island (being in the charge state $k$) through $i$th junction, $R_T/2$ is the tunnel resistance of each junction.
Here we focus on the magnetic field effects in the sample B (see Fig.~2(b) in the main text) and neglect all the overheating effects assuming the equilibrium Fermi distribution of electrons over energy $f_T(E) = [e^{E/k_{\rm B} T}+1]^{-1}$ with the electron temperature $T$ equal to the phonon bath temperature $T_0$.
The density of states (DOS) $n_S(E)={\rm Re} [\cos\theta]$ normalized to its normal state value $D(E_F)$ in the superconducting (S) island near the junction is obtained from the solution of averaged Usadel equation, Eq.~(1) from the main text, with the depairing parameter $\Gamma/\Delta_0=\alpha_1 \left(B/B_c\right)^2 - m\alpha_2 B/B_c +m^2\alpha_3$, Eq.~(2) in the main text, having three positive numerical fitting parameters $\alpha_l$.

In the Sample B the tails of the wave functions localized in the vortex core(s) give a substantial contribution to the DOS and to $IV$ curves for $m\ne 0$ at small $I_{\rm bias}$, but they are not included into the averaged model.
To model this contribution we replace the DOS $n_S(E)$ by $n_S(E)(1-e^{-R/r_v})+e^{-R/r_v}$ by adding the phenomenological normal metal DOS with the exponentially suppressed prefactor $e^{-R/r_v}$ determined by the vortex distance from the junction $R\sim 0.5$~$\mu$m and by the exponential decay of the wave function localized in the vortex core of the effective radius $r_v$.
This vortex contribution leads to reduced $V(B)$ at small $I_{\rm bias}$ in the mixed state $m\ne 0$ and to the suppression of the jumps at the vortex entry fields (see red and green curves in Supplementary Figure~\ref{fig1SM_dc_V_vs_B}).

By fitting $V(B)$ at $I_{\rm bias}=100$~pA which is not affected by the vortex tail contributions one can extract the following values of fitting parameters $\alpha_1=0.38$, $\alpha_2=0.438$, and $\alpha_3=0.266$ mentioned in the main text.
Following \cite{Kanda2004} we attribute to all jumps in this plot with the change of the number of vortices in the sample and use the point of the first jump at $B>0$ as the field of the first vortex entry $B_c=14.4$~mT. In this setup we don't see any transitions between vortex configurations with the constant vorticity like the transition to a giant vortex state (see, e.g., \cite{Schweigert1998, Palacios1998}).
Using these parameters one can fit $V(B)$ quite well at all bias current values with $R/r_v\sim 1.7$.  The optimal value of the vortex core radius $r_v=2.5-2.7\xi$ extracted from dc measurements in the Sample B is in perfect agreement with the previous theoretical works \cite{Golubov1993,Golubov1994}.
In subgap regime $I_{\rm bias}=1$ and $10$~pA the jump-like anomalies in $V(B)$ become knee-like, but because of the above-mentioned reasons we still associate each of them with the vortex entry or exit.

\section*{Supplementary Note 4. Electronic pumping}\label{SM_sec5:pumping}

The electronic pumping of the Sample B at $f=5$~MHz, when $V_{\rm bias}\approx 120~\mu$V and $n_g\sim0.5$, is shown in Supplementary Figure~\ref{fig3SM_Pumping}(a) with the field $H$ swept from $-10$~mT to 2~mT. Contrary to what is observed in the Sample A, the increase of the magnetic field increases the deviation from the current quantization $I=ef$, due to the effect of the screening current on the superconducting gap.
This observation is in agreement with the theoretical model with the increasing number of QPs in S island with the field.

\begin{figure}[h!]
	\centering{
		\includegraphics[width=0.96\textwidth]{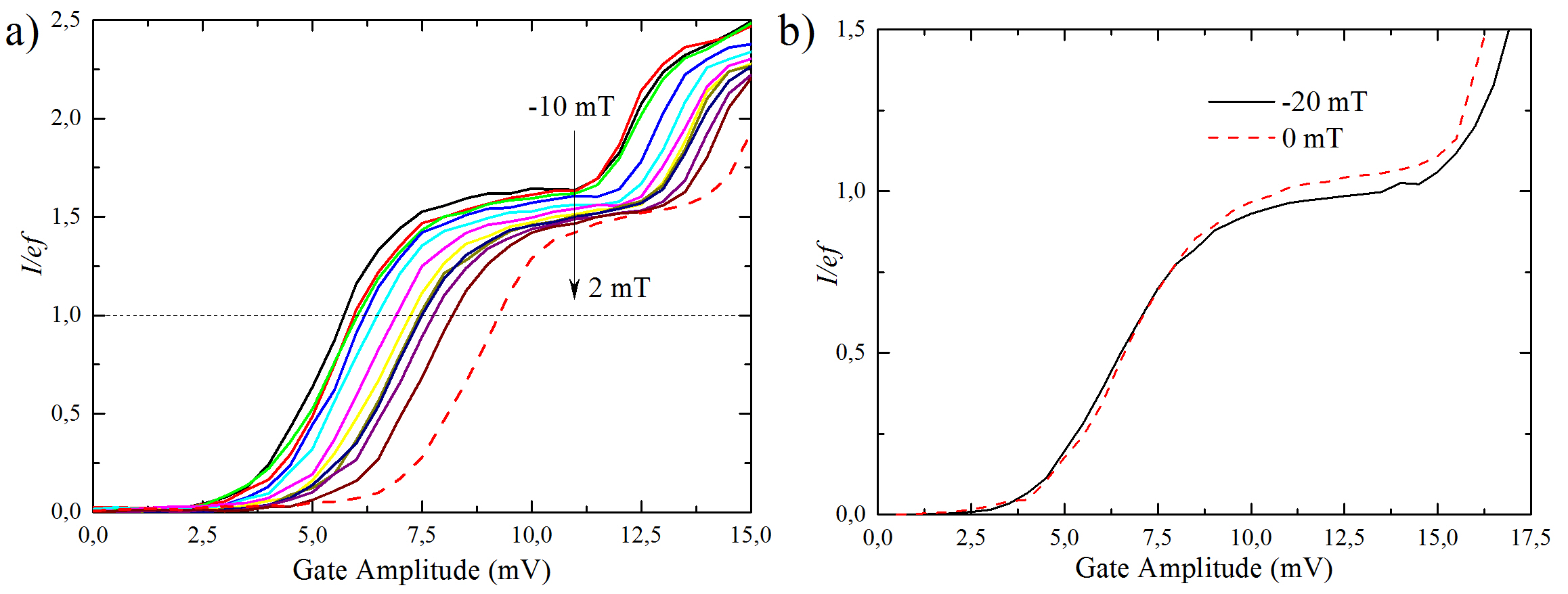}
    }
	\caption{{\bf Pumping of Sample B and sample A at 200\,MHz.} (a) Electronic pumping of the Sample B at $f=5$~MHz at fixed bias voltage $V_{\rm bias}=120$~$\mu$V under field $H$ from $-10$~mT to $2$~mT (from top to bottom). With the field, the deviation from $I=ef$ increases rather than decreases in contrast to the behavior of Sample A.
(b) Electronic pumping of the Sample A at $f=200$~MHz at fixed bias voltage $V_{\rm bias}=250$~$\mu$V in field $H=0$ (red dashed line) and $H=-20$~mT (solid black line).}
	\label{fig3SM_Pumping}
\end{figure}

The electronic pumping of the Sample A at frequency $f=200$~MHz, when $V_{\rm bias}\approx 250~\mu$V and $n_g\sim0.5$, is shown in Supplementary Figure~\ref{fig3SM_Pumping}(b) for two field values $H=0$ and $-20$~mT. Similarly to the lower frequency range the increase of the magnetic field reduces the QP contribution to the excess current.

The evolution of the pumping current in the Sample A with the field is shown in Supplementary Figure~\ref{fig4SM_dI_vs_H_diff_range+large}.
In panel (a) we show the pumping current versus field with the different initial field values.
For a small value of the initial field $H=-12$~mT, the island is in the Meissner state and a continuous variation of current is observed.
The anomaly at $-2$~mT appears only if the initial field is large enough to turn island into the mixed state (see blue and red curves).
Similarly, the anomaly at $H_{out}^{(2)}\sim-15$~mT (see inset of Fig. 4 of main text) appears only if the initial field exceeds $20$~mT.
Panel (b) shows that the current deviates significantly from $I=ef$ at fields larger than $\sim30$~mT due to the reduction of the S gap near the junctions.

\begin{figure}
	\centering
		\includegraphics[width=0.96\textwidth]{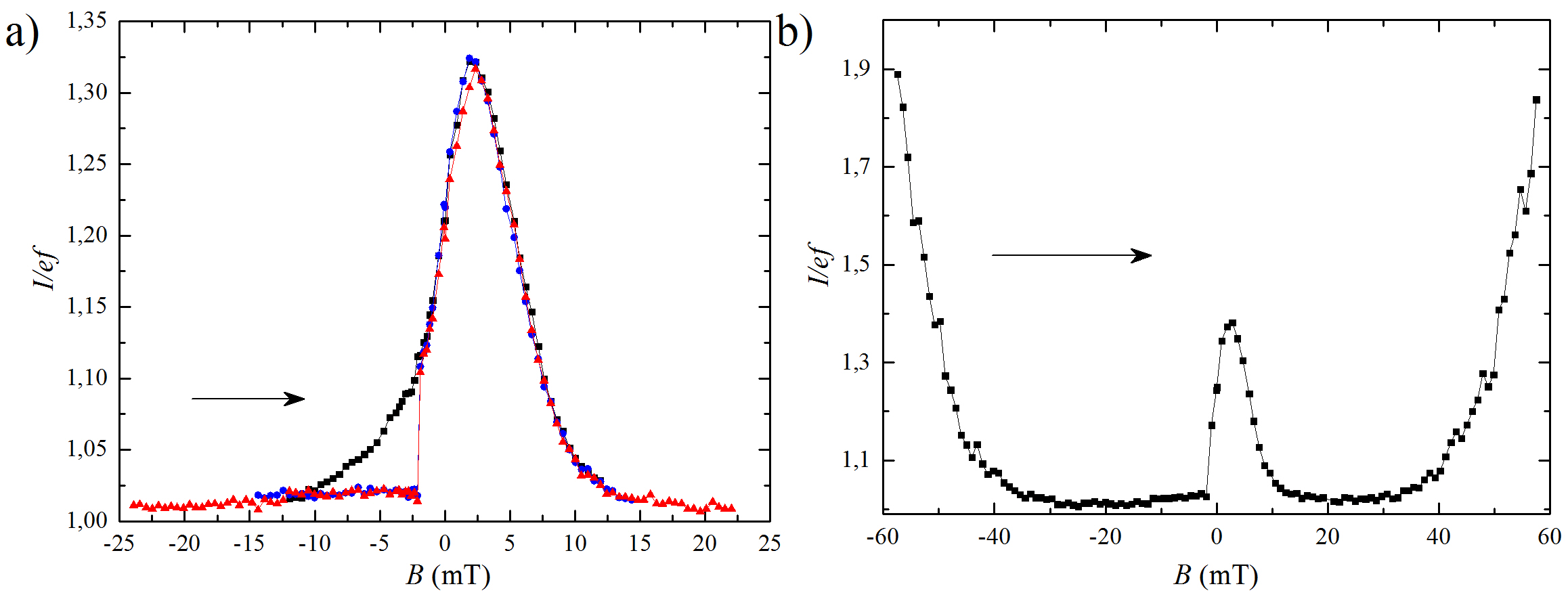}
	\caption{{\bf Extra pumping measurements of Sample A.} (a) Evolution of the electronic pumping with the magnetic field measured in the range $\pm12$~mT (black squares), $\pm15$~mT (blue open cirles) and $\pm25$~mT (red crosses). At the initial field value the island is in a vortex free, in a single-vortex, or in a two-vortex state, respectively. (b) Pumping current in a wide field range, from $-60$~mT to 60~mT.
The pumping has been measured with the gate offset $n_g^0=0.5$ and the bias voltage $V_{\rm bias}\simeq 100$~$\mu$V at $f=5$~MHz.}
	\label{fig4SM_dI_vs_H_diff_range+large}
\end{figure}

\section*{Supplementary Note 5. Heat balance equation}\label{SM_sec6:heat_balance}
In this section we describe the theoretical model of the relaxation of QPs in applied magnetic field by using the example of NISIN SET
in the turnstile regime.
By applying the constant bias voltages $V_{L,R}=\pm V_{\rm bias}/2$ to the normal leads and the periodic gate voltage $n_g(t) = C_g V_g(t)/e = n_g^0+A_g \sin(2\pi f t)$ with a certain offset $n_g^{0}$, frequency $f$, and the amplitude $A_g$ to the gate electrode 
one can push electrons to tunnel through the system producing a time-dependent current $I(t)$.
This transport current $I(t)$ flowing from one lead to another drives the NISIN turnstile out of the equilibrium by injecting nonequilibrium QPs into the S island.
The power $\dot{Q}_{NIS}^S$ injected to the island increases with the frequency $f$ and we model this increase in mean density of QPs in the superconductor by raising its electron temperature $T$ relatively to the phonon bath temperature $T_0$.
Note that the quasiequilibrium Fermi distribution of electrons over energy $f_T(E) = [e^{E/k_{\rm B} T}+1]^{-1}$ is provided by the smallness of the inelastic electron-electron scattering time $\tau_{ee}$ comparing to the operating time $\tau_0=1/f$ and the effective charging time $e/I$.\cite{Giazotto2006}

Due to the large electron-phonon relaxation length $L_T\gg R$ compared to the island size $R$ we consider the heat balance equation \cite{Peltonen2011} for the whole island
\begin{gather}
\dot{Q}_{NIS}^S = \dot{Q}_{eph} \ ,
\end{gather}
where $\dot{Q}_{eph}$ is the electron-phonon heat flow averaged over the island volume $\mathcal{V}$.
The power injected into the island from the junction can be written as follows
\begin{gather}\label{Q_NIS^S}
\dot{Q}_{NIS}^S = \sum_{k,i} p_k\left[\dot{Q}_S(U_{k,i}^+)+\dot{Q}_S(U_{k,i}^-)\right]
\end{gather}
as a sum of the power dissipated in the S island in a single tunneling event
\begin{gather}\label{Q_S(U)}
\dot{Q}_S(U)=\frac{1}{e^2 R_T}\int E_S n_S(E)f_T(E)\left[1-f_{T_0}(E+U)\right]dE \ ,
\end{gather}
over the junction $i=L,R$ through which electron tunnels and over the island charge state $k$. This sum is weighted with the probability $p_k$ of system being in this charge state, which is calculated using the standard rate equation for the balance of the probability fluxes Sup. Eq.~\eqref{master_eq_k}.
Here 
$E_S=E$.
Note that here and further on we neglect the difference between the electronic temperature $T_N$ in the normal metal of volume $\mathcal{V}_N$ and the bath temperature $T_0$, because of sufficient electron-phonon relaxation there $Q_{eph}^{N} = \Sigma_N \mathcal{V}_N (T_N^5 - T_0^5)$ ($\Sigma_N$ is the electron-phonon material constant) and the rather small power $\dot{Q}_{NIS}^N$ injected into the normal leads which can be obtained from Sup. Eq.~\eqref{Q_NIS^S} by replacing $E_S$ by $E_N=eV_i-E$ in Sup. Eq.~\eqref{Q_S(U)}.

Due to the Joule's law the sum of the powers injected into the island $\dot{Q}_{NIS}^S$ and into the normal leads $\dot{Q}_{NIS}^N$ should be equal to $I V$, where $I$ is the current in Sup. Eq.~\eqref{I_pk} averaged over the period $\tau_0=1/f$ of the gate voltage operation.
Usually if the superconductor is not overheated much $k_{\rm B} T\ll E_g$ one can assume that $\dot{Q}_{NIS}^S\gg \dot{Q}_{NIS}^N$ in subgap regime and $\dot{Q}_{NIS}^S\simeq \dot{Q}_{NIS}^N$ at biases above the gap \cite{Giazotto2006}.
As a result within constant factor of order of unity the heat balance equation can be rewritten as follows
\begin{gather}\label{Heat_balance_eq}
\dot{Q}_{eph} \simeq I V \ .
\end{gather}
Within the optimal conditions of the proper turnstile shielding and optimized device geometry the averaged current $I$ is close to its ideal value $e f$ and the deviation $\delta I=I - e f$ is mainly governed by nonequilibrium QP density in the S island near the junction
\begin{gather}\label{n_qp}
n_{qp}(T)=2 D(E_F)\int_0^\infty n_S( E)f_T(E) dE \approx D(E_F)\sqrt{2\pi k_{\rm B} T \Delta_0} e^{-\Delta_0/k_{\rm B} T} \ .
\end{gather}
The estimates for contributions in higher orders in small parameter $\hbar/e^2 R_T$ are given in the following Note.
The latter expression is written for the sample A where the S gap near the junction is close to $\Delta_0$.
At $B=0$ the QP number $N_{qp}$ in the S island equals $N_{qp}=n_{qp}\mathcal{V}$ the product of $n_{qp}$ to the volume of the island $\mathcal{V}$.
The QP number $N_{qp,vort}\simeq 2 D(E_F) \mathcal{V}_v k_{\rm B} T \ln 2$ in the vortex core can be obtained from Sup. Eq.~\eqref{n_qp} by substitution of the normal state DOS $n_S(E)=1$ into the integral and by multiplying it by  the vortex core volume $\mathcal{V}_v = r_v^2 d_S\sim 7\xi^2 d_S$.

The electron-phonon heat flux under magnetic field is similar to the expression given by Eq.~(3) in \cite{Maisi2013}
\begin{gather}\label{Q_eph}
\dot Q_{eph}=\frac{\Sigma \mathcal{V}}{24 \zeta(5) k_{\rm B}^5}\int_{0}^{\infty}\epsilon^3\left[n_{T}(\epsilon)-n_{T_0}(\epsilon)\right] d\epsilon \int_{-\infty}^{\infty} M_{E,E+\epsilon}\left[f(E)-f(E+\epsilon)\right]dE \ .
\end{gather}
with the term $n_S(E)n_S(E+\epsilon)[1-\Delta_0^2/E(E+\epsilon)]$ substituted by $M_{E,E+\epsilon}$ of the form \cite{Kopnin2001}
\begin{gather}\label{M_E,E'_def}
8M_{E,E'}=2(g^R_{E'}-g^A_{E'})(g^R_{E}-g^A_{E})-(f^R_{E'}-f^A_{E'})(f^{\dagger R}_{E}-f^{\dagger A}_{E})-(f^R_{E}-f^A_{E})(f^{\dagger R}_{E'}-f^{\dagger A}_{E'}) \ .
\end{gather}
Here $\Sigma$ is the electron-phonon material constant, and $\zeta(s)$ is the Riemann zeta function. The retarded (advanced) normal $g^{R(A)}$ and anomalous $f^{R(A)}$ Green's functions are determined by the solution of Eq.~(1) from the main text, i.e., $M_{E,E'} = n_S(E)n_S(E')-b(E)b(E')$, with $n_S(E)={\rm Re} [\cos\theta]$ and $b(E)={\rm Im}[\sin\theta]$.

In the low temperature limit $T_0,T\ll E_g/k_{\rm B}$ the main contribution to Sup. Eq.~\eqref{Q_eph} arises from the energies $0<|E|-E_g\lesssim k_{\rm B} T$ close to the hard gap value $\pm E_g$, which can be calculated using the following expansion of $n_S(E)$ and $b(E)$ over the small positive parameter $\delta E=|E|-E_g\ll E_g$
\begin{gather}\label{DOS}
n_S(E)^2 = \Theta(\delta E)\frac{2\delta E\Delta^{2/3}}{3\Gamma^{4/3}E_g^{1/3}}\ , \quad \frac{b(E)}{n_S(E){\rm sign}(E)} \approx \left(\frac{E_g}{\Delta}\right)^{1/3} \ .
\end{gather}
Here $\Theta(x)$ is the Heaviside theta-function.

Substituting Sup. Eqs.~(\ref{M_E,E'_def}, \ref{DOS}) into Sup. Eq.~\eqref{Q_eph} and taking into account only the leading terms in the small parameter $k_{\rm B} T/E_g$ we obtain
\begin{gather}\label{Q_eph_low_T_zero_H}
\dot Q_{eph}=\frac{\Sigma \mathcal{V}}{\zeta(5)}\left\{ \frac{64}{63} T^5 e^{-\frac{E_g}{k_{\rm B} T}}
+ \frac{2\pi E_g^4}{3k_{\rm B}^4} T e^{-\frac{2E_g}{k_{\rm B} T}}\right\}
\end{gather}
for rather large electronic temperatures $\Gamma^{2/3} E_g^{1/3}/k_{\rm B}, T_0\ll T\ll E_g/k_{\rm B}$.
Note that the recombination term ($\propto e^{-{2E_g}/{k_{\rm B} T}}$) dominates at $k_{\rm B}T>0.1 E_g$ and should be taken into account.
In the opposite case $T_0\ll T\ll \Gamma^{2/3} E_g^{1/3}/k_{\rm B}$
\begin{gather}\label{Q_eph_low_T_in_H}
\dot Q_{eph}=\frac{\Sigma \mathcal{V} T^3}{9\zeta(5)\Gamma^{2/3} E_g^{1/3}}\left\{ \frac{128}{21} k_{\rm B} T^3 e^{-\frac{E_g}{k_{\rm B} T}}
+ \frac{\pi E_g^3}{k_{\rm B}^2} e^{-\frac{2E_g}{k_{\rm B} T}}\right\}
\end{gather}
Here the recombination term is of order of the scattering term ($\propto e^{-{E_g}/{k_{\rm B} T}}$) at $k_{\rm B}T\sim 0.3E_g$.
In both cases as the temperature becomes of the order of the gap one have to use full numerical expression given by Sup. Eq.~\eqref{Q_eph}.

\section*{Supplementary Note 6. Excess current as a function of electronic temperature}\label{SM_sec7:dI}
To calculate our main observable, the leakage current $\delta I=I - e f$ in the NISIN turnstile we use the simplified version of the master equation given in Sup. Eq.~\eqref{master_eq_k} for low temperatures taking into account only two charge states $k=0$ and $k=1$
\begin{gather}
\frac{d p_1}{dt} = \Gamma_{0\to 1} p_{0}-\Gamma_{1\to 0} p_1 \ , \quad p_0 = 1-p_1 \ .
\end{gather}
with the tunneling rates given in Sup. Eq.~\eqref{Gamma[U]} in the subgap regime $|U_{0,i}^+|<E_g$ given by
\begin{gather}\label{Gamma[U]_subgap}
\Gamma[U]\approx \Gamma_{T_0}e^{-(E_g-U)/k_{\rm B} T_0}+\Gamma_{T}e^{-E_g/k_{\rm B} T} \
\end{gather}
Here $U_{0,i}^+=-U_{1,i}^-=2E_C A_g \sin(2\pi f t)-eV_i$ and $\Gamma_T=\sqrt{2\pi k_{\rm B} T E_g}/e^2 R_T$.
This expression contains the exponentially growing part with $U$ which determines the dominant tunneling rate with maximal $U$ for each time instant.

We consider the offset $n_g^0=0.5$ for simplicity
and use the symmetry of the drive $n_g(\tau_0-t)=1-n_g(t)$ focusing on the first half of the period with $n_g$ increasing from $0.5-A_g$ to $0.5+A_g$.
We assume that before the time instant $t_1$ the island is discharged $k=0$ due to the domination of rate $\Gamma_{1\to 0}^{R}$ among the others and the charging process is started at $t=t_1$.
The probability $P_0(t)$ to stay in the state $k=0$ is decreasing with time $t>t_1$ as
\begin{gather}\label{P_0_charging}
P_0(t) =  \exp\left[-\int_{t_1}^{t}\Gamma_{0\to 1}(t') dt'\right] .
\end{gather}
For typical frequencies the charging process occurs not far from $n_g = 0.5$, therefore further we linearize the drive $n_g(t)\approx 0.5 + 2 \pi A_g (f t-1/4)$.
As the island has been charged $P_0(t^*)=\epsilon\lesssim 1$ (let's take $\epsilon=1/2$ for definiteness) the leakage current starts to flow.
The number of excess electrons $N_l$ through the island can be written as the integral of the largest subleading rate $\Gamma_{1\to 0}^R(t)$ governing the leakage current over the time interval $t^*<t<t_2$ before this rate becomes the dominant one
\begin{gather}\label{excess_N_l}
N_l\simeq \int_{t^*}^{t_2} \Gamma_{1\to 0}^R(t) dt \ .
\end{gather}
The leakage current can be calculated as follows $\delta I\simeq 2 e f N_l$, where '2' accounts for the leakage during the second half of the period due to the symmetry $k\leftrightarrow 1-k$ and $L\leftrightarrow R$.

By substituting Sup. Eq.~\eqref{Gamma[U]_subgap} in Sup. Eqs.~(\ref{P_0_charging}, \ref{excess_N_l}) and calculating the time instants $t_1=\tau_0-t_2$ and $t_*$ one can come to the result
\begin{gather}\label{dI}
\delta I = I-ef \simeq  e\Gamma_{T}\left[1-\frac{2 E_g-|e|V - k_{\rm B} T_0 a(T)}{2\pi E_C A_g}\right]e^{-\frac{E_g}{k_{\rm B} T}} \ ,
\end{gather}
where $a=\ln\frac{\Gamma_{T_0}k_B T_0}{2\pi E_C A_g f\ln 2}$ for $n_g(t_1)=n_g(t_2)=0.5$ at low enough electronic temperature $T\lesssim T_0[1-(eV/2E_g)]^{-1}$ and $a(T)=|e|V/2-\frac{E_g}{k_B} (1-T_0/T)+\ln\frac{\Gamma_{T_0}k_B T_0^2}{2\pi E_C A_g f T\ln 2}$ for $n_g(t_1)=1-n_g(t_2)<0.5$ in the opposite case $T\gtrsim T_0[1-(eV/2E_g)]^{-1}$.
In this derivation we consider the operating frequency $f$ to be small compared to the charging rate $\sim(t_*-t_1)^{-1}$ to avoid missing events. 
We neglect the relative corrections of order of $e^{-|e|V/k_{\rm B} T_0}$ ($\Gamma_{0\to 1}^R/\Gamma_{0\to 1}^L$ and $\Gamma_{1\to 0}^L/\Gamma_{1\to 0}^R$ for the case when the first term in Sup. Eq.~\eqref{Gamma[U]_subgap} dominates for all rates).
We don't take into account the factor $1/2$ in $N_l$ during the time when $\Gamma_{0\to 1}^L\gg\Gamma_{1\to 0}^L=\Gamma_{1\to 0}^R\simeq \Gamma_{T}e^{-E_g/k_{\rm B} T}$, when the discharging occurs with the equal probability $p_{L,R}=1/2$ to the left and to the right contact.
We can do it, because during the integration of Sup. Eq.~\eqref{excess_N_l} $U_{1,L}^{-}$ can go beyond the subgap range $U_{1,L}^{-}<-E_g$ suppressing the second term in Sup. Eq.~\eqref{Gamma[U]_subgap} for $\Gamma_{1\to 0}^L$ exponentially $\sim e^{-(E_g+|U_{1,L}^{-}|)/k_{\rm B} T}$ and keeping the rate $\Gamma_{1\to 0}^R$ to be the dominant one in the leakage current.

To avoid all these unimportant details we consider a certain $A_g$-dependent numerical prefactor $C\sim 1$ instead of the square brackets in Sup. Eq.~\eqref{dI} and come to Eq.~(5) of the main text by using the assumption that the S gap near the junction (in the sample A) is close to $\Delta_0$.

Comparing Sup. Eq.~\eqref{n_qp} and Eq.~(5) in the main text one can write down the following relation between the leakage current $\delta I = I - e f$ and the QP density $n_{qp}$ near the junctions
\begin{gather}
n_{qp} = D(E_F) e R_T \delta I / C
\end{gather}
used in Fig.~4 of the main text to show the QP density scale.

Note that we neglect also the contributions of higher orders in the small parameter $\hbar/(e^2 R_T)$ like Andreev tunneling (see, e.g., \cite{Aref2011,Averin2008}) due to rather large tunnel resistance of the sample contacts.
Indeed, from the experimental side the attribute feature of Andreev tunneling is the additional peak in the beginning of each current plateau $I=n e f$ \cite{Aref2011} which is not observed in all pumping measurements of this paper.
From the theoretical side one can estimate the relative contribution $\delta I_{AR}/(ef)$ of Andreev tunneling to the current as the ratio $\Gamma_{AR}/\Gamma[U]$ of dc rates of sequential $\Gamma[U]=U/(e^2 R_T)$ and Andreev tunneling $\Gamma_{AR}\simeq \pi \hbar U/(4 N e^4 R_T^2)$ in the above-gap regime.
Here $N=A/A_{ch}$ is the number of channels in the tunnel junction, $A\simeq 6\cdot 10^3$~nm$^2$ is the area of the junction and $A_{ch}$ is the are of a single channel. Theoretical estimates given in \cite{Averin2008} lead to $A_{ch}\sim 2$~nm$^2$, while experimental observation \cite{Aref2011} gives $A_{ch}\sim 30$~nm$^2$. The upper bound estimate with $A_{ch}\sim 30$~nm$^2$ and $R_T=577$~k$\Omega$ for the sample A gives $N\sim 200$ and $\delta I_{AR}/(ef)\sim\pi \hbar /(4 N e^2 R_T)\simeq 3\cdot 10^{-5}$ which can be neglected comparing to the QP contribution.



\begin{thebibliography}{35}%
\subsection*{References}
\bibitem {Martinis2009}	
	Martinis, J. M., Ansmann, M. \& Aumentado, J. Energy decay in superconducting josephson-junction qubits from nonequilibrium quasiparticle excitations. \textit{Phys. Rev. Lett.} \textbf{103,} 097002 (2009).
\bibitem {Paik2011}
	Paik, H. \textit{et al.} Observation of high coherence in josephson junction qubits measured in a three-dimensional circuit qed architecture. \textit{Phys. Rev. Lett.} \textbf{107,} 240501 (2011).
\bibitem {Corcoles2011}%
	C\`orcoles, A. D. \textit{et al.} Protecting superconducting qubits from radiation. \textit{Appl. Phys.  Lett.} \textbf{99,} 181906 (2011).
\bibitem {Wang2009}
	Wang, H. \textit{et al.} Improving the coherence time of superconducting coplanar resonators. \textit{Appl. Phys. Lett.} \textbf{95,} 233508 (2009).
\bibitem {Barends2011}%
	Barends, R. \textit{et al.} Minimizing quasiparticle generation from stray infrared light in superconducting quantum circuits. \textit{Appl. Phys. Lett.} \textbf{99,} 113507 (2011).
\bibitem {Knowles2012}%
	Knowles, H. S., Maisi, V. F. \& Pekola, J. P. Probing quasiparticle excitations in a hybrid single electron transistor. \textit{Appl. Phys.  Lett.} \textbf{100,} 262601 (2012).
\bibitem {Pekola2000}
	Pekola, J. P. \textit{et al.} Trapping of quasiparticles of a nonequilibrium superconductor. \textit{Appl. Phys. Lett.} \textbf{76,} 2782 (2000).
\bibitem {Rajauria2009}
	Rajauria, S., Courtois, H. \& Pannetier, B. Quasiparticle-diffusion-based heating in superconductor tunneling microcoolers. \textit{Phys. Rev. B} \textbf{80,} 214521 (2009).
\bibitem {Nguyen2013}	
	Nguyen, H. Q. \textit{et al.} Trapping hot quasi-particles in a high-power superconducting electronic cooler. \textit{New Journal of Physics} \textbf{15,} 085013 (2013).
\bibitem {Ullom1998}
	Ullom, J. N., Fisher, P. A. \& Nahum, M. Magnetic field dependence of quasiparticle losses in a superconductor. \textit{Appl. Phys. Lett.} \textbf{73,} 2494-2496 (1998).
\bibitem {Yamamoto2006}
	Yamamoto, T., Nakamura, Y., Pashkin, Y. A., Astafiev, O. \& Tsai, J. S. Parity effect in superconducting aluminum single electron transistors with spatial gap profile controlled by film thickness. \textit{Appl. Phys. Lett.} \textbf{88,} 212509 (2006).
\bibitem{Chi1979}
    Chi, C. C. \& Clarke, J. Enhancement of the energy gap in superconducting aluminum by tunneling extraction of quasiparticles. \textit{Phys. Rev. B} \textbf{20,} 4465-4473 (1979).
\bibitem{Blamire1991}
    Blamire, M. G., Kirk, E. C. G., Evetts, J. E. \& Klapwijk, T. M. Extreme Critical-Temperature Enhancement of Al by Tunneling in Nb/AlOx/Al/AlOx/Nb Tunnel Junctions. \textit{Phys. Rev. Lett.} \textbf{66,} 220-223 (1991).
\bibitem{Heslinga1993}
    Heslinga, D. R. \& Klapwijk, T. M. Enhancement of superconductivity far above the critical temperature in double-barrier tunnel junctions. \textit{Phys. Rev. B} \textbf{47,} 5157-5164 (1993).
\bibitem {Goldie1990}%
	Goldie, D. J., Booth, N. E., Patel, C. \& Salmon, G. L. Quasiparticle trapping from a single-crystal superconductor into a normal-metal film via the proximity effect. \textit{Phys. Rev. Lett.} \textbf{64,} 954-957 (1990).
\bibitem {Ullom2000}
	Ullom, J. N., Fisher, P. A. \& Nahum, M. Measurements of quasiparticle thermalization in a normal metal. \textit{Phys. Rev. B} \textbf{61,} 14839-14843 (2000).
\bibitem {Rajauria2012}
	Rajauria, S. \textit{et al.} Efficiency	of quasiparticle evacuation in superconducting devices.	\textit{Phys. Rev. B} \textbf{85,} 020505(R) (2012).
\bibitem {Levenson-Falk2014}%
	Levenson-Falk, E. M., Kos, F., Vijay, R., Glazman, L. \& Siddiqi, I. Single-quasiparticle trapping in aluminum nanobridge josephson junctions. \textit{Phys. Rev. Lett.} \textbf{112,} 047002 (2014).
\bibitem {Golubov1993}%
	Golubov, A. A. \& Houwman, E. P. Quasiparticle relaxation rates in a spatially inhomogeneous superconductor. \textit{Physica C: Superconductivity} \textbf{205,} 147-153 (1993).
\bibitem {Golubov1994}%
	Golubov, A. A. \textit{et al.} Quasiparticle lifetimes and tunneling times in a superconductor-insulator-superconductor tunnel junction with spatially inhomogeneous electrodes. \textit{Phys. Rev. B} \textbf{49,} 12953-12968 (1994).
\bibitem {Friedrich1997}%
	Friedrich, S. \textit{et al.} Experimental quasiparticle dynamics in a superconducting, imaging x-ray spectrometer. \textit{Appl. Phys.  Lett.} \textbf{71,} 3901-3903 (1997).
\bibitem {Aumentado2004}%
	Aumentado, J., Keller, M. W., Martinis, J. M. \& Devoret, M. H. Nonequilibrium quasiparticles and 2e periodicity in single-cooper-pair transistors. \textit{Phys. Rev. Lett.} \textbf{92,} 066802 (2004).
\bibitem {Court2008}%
	Court, N. A., Ferguson, A. J., Lutchyn, R. \& Clark, R. G. Quantitative study of quasiparticle traps using the single-Cooper-pair transistor. \textit{Phys. Rev. B} \textbf{77,} 100501(R) (2008).
\bibitem {Peltonen2011}
	Peltonen, J. T., Muhonen, J. T., Meschke, M., Kopnin, N. B. \& Pekola, J. P. Magnetic-field-induced stabilization of nonequilibrium superconductivity in a normal-metal/insulator/superconductor junction. \textit{Phys. Rev. B} \textbf{84,} 220502(R) (2011).
\bibitem {Nsanzineza2014}
	Nsanzineza, I. \& Plourde, B. L. T. Trapping a single vortex and reducing quasiparticles in a superconducting resonator. \textit{Phys. Rev. Lett.} \textbf{113,} 117002 (2014).
\bibitem {Wang2014}
	Wang, C. \textit{et al.} Measurement and control of quasiparticle dynamics in a superconducting qubit. \textit{Nature Commun.} \textbf{5,} 5836 (2014).
\bibitem {Woerkom2015}
	van Woerkom, D. J., Geresdi, A. \& Kouwenhoven, L. P. One minute parity lifetime of a NbTiN Cooper-pair transistor. \textit{Nature Phys.}, \textbf{11}, 547–550 (2015).
\bibitem{Frungel2014}
    Fr\"ungel, F. B. A. \textit{Optical Pulses - Lasers - Measuring Techniques}. (Academic Press, New York, 2014).
\bibitem {Pekola2008}
	Pekola, J. P. \textit{et al.} Hybrid single-electron transistor as a source of quantized electric current. \textit{Nature Phys.} \textbf{4,} 120-124 (2008).
\bibitem {Kanda2004}%
	Kanda, A., Baelus, B. J., Peeters, F. M., Kadowaki, K. \& Ootuka, Y. (2004). Experimental evidence for giant vortex states in a mesoscopic superconducting disk. \textit{Phys. Rev. Lett.} \textbf{93,} 257002 (2004).
\bibitem {Stan2004}
	Stan, G., Field, S. B. \& Martinis, J. M. Critical field for complete vortex expulsion from narrow superconducting strips. \textit{Phys. Rev. Lett.} \textbf{92,} 097003 (2004).
\bibitem {Skalski1964}
	Skalski, S., Betbeder-Matibet, O. \& Weiss, P. R. Properties of superconducting alloys containing paramagnetic impurities. \textit{Phys. Rev.} \textbf{136,} A1500-A1518 (1964).
\bibitem {Maki1965}	
	Maki, K. \& Fulde, P. Equivalence of different pair-breaking mechanisms in superconductors. \textit{Phys. Rev.} \textbf{140,} A1586-A1592 (1965).
\bibitem {Fulde1965}%
	Fulde, P. Tunneling density of states for a superconductor carrying a current. \textit{Phys. Rev.} \textbf{137,} A783-A787 (1965).
\bibitem {Anthore2003}%
	Anthore, A., Pothier, H. \& Esteve, D. Density of states in a superconductor carrying a supercurrent. \textit{Phys. Rev. Lett.} \textbf{90,} 127001 (2003).
\bibitem{Schweigert1998}%
  Schweigert, V. A., Peeters, F. M. \& Singha Deo, P. Vortex Phase Diagram for Mesoscopic Superconducting Disks. \textit{Phys. Rev. Lett.} \textbf{81,} 2783-2786 (1998).
\bibitem{Palacios1998}%
  Palacios, J. J. Vortex matter in superconducting mesoscopic disks: Structure, magnetization, and phase transitions. \textit{Phys. Rev. B} \textbf{58,} R5948-R5951(R) (1998).
\bibitem {Kopnin2001}%
	Kopnin, N. B. \textit{Theory of Nonequilibrium Superconductivity}. (Oxford Univ. Press, Oxford, 2001).
\bibitem {Timofeev2009}
	Timofeev, A. V. \textit{et al.} Recombination-Limited Energy Relaxation in a Bardeen-Cooper-Schrieffer Superconductor. \textit{Phys. Rev. Lett.} \textbf{102,} 017003 (2009).
\bibitem {Maisi2013}	
	Maisi, V. F. \textit{et al.} Excitation of Single Quasiparticles in a Small Superconducting Al Island Connected to Normal-Metal Leads by Tunnel Junctions. \textit{Phys. Rev. Lett.} \textbf{111,} 147001 (2013).
\bibitem{Aref2011}%
  Aref, T., Maisi, V. F., Gustafsson, M. V., Delsing P., \& Pekola, J. P. Andreev tunneling in charge pumping with SINIS turnstiles. \textit{Europhys. Lett.} \textbf{96,} 37008 (2011).
\bibitem{Kautz1993}
   Kautz, R. L., Zimmerli, G. \& Martinis, J. M. Self-heating in the Coulomb-blockade electrometer.
   \textit{Jour. Appl. Phys.} \textbf{73,} 2386-2396 (1993).
\bibitem {Giazotto2006}%
	Giazotto, F., Heikkil\"a, T. T., Luukanen, A., Savin, A. M. \& Pekola, J. P. Opportunities for mesoscopics in thermometry and refrigeration: Physics and applications. \textit{Rev. Mod. Phys.} \textbf{78,} 217-274 (2006).
\end{thebibliography}

\begin{thebibliography}{99}%
\subsection*{Supplementary References}
\bibitem {Anthore2003}%
	Anthore, A., Pothier, H. \& Esteve, D. Density of states in a superconductor carrying a supercurrent. \textit{Phys. Rev. Lett.} \textbf{90,} 127001 (2003).
\bibitem {Golubov1993}%
	Golubov, A. A. \& Houwman, E. P. Quasiparticle relaxation rates in a spatially inhomogeneous superconductor. \textit{Physica C: Superconductivity} \textbf{205,} 147-153 (1993).
\bibitem {Golubov1994}%
	Golubov, A. A. \textit{et al.} Quasiparticle lifetimes and tunneling times in a superconductor-insulator-superconductor tunnel junction with spatially inhomogeneous electrodes. \textit{Phys. Rev. B} \textbf{49,} 12953-12968 (1994).
\bibitem {Peltonen2011}
	Peltonen, J. T., Muhonen, J. T., Meschke, M., Kopnin, N. B. \& Pekola, J. P. Magnetic-field-induced stabilization of nonequilibrium superconductivity in a normal-metal/insulator/superconductor junction. \textit{Phys. Rev. B} \textbf{84,} 220502(R) (2011).
\bibitem {Skalski1964}
	Skalski, S., Betbeder-Matibet, O. \& Weiss, P. R. Properties of superconducting alloys containing paramagnetic impurities. \textit{Phys. Rev.} \textbf{136,} A1500-A1518 (1964).
\bibitem {Maki1965}	
	Maki, K. \& Fulde, P. Equivalence of different pair-breaking mechanisms in superconductors. \textit{Phys. Rev.} \textbf{140,} A1586-A1592 (1965).
\bibitem {Fulde1965}%
	Fulde, P. Tunneling density of states for a superconductor carrying a current. \textit{Phys. Rev.} \textbf{137,} A783-A787 (1965).
\bibitem {Averin1986}
    Averin, D. \& Likharev, K.  Coulomb blockade of single-electron tunneling, and coherent oscillations in small tunnel junctions. \textit{Journal of Low Temperature Physics} \textbf{62,} 345-373 (1986).
\bibitem {Fulton1987}	
    Fulton, T. A. \& Dolan, G. J. Observation of single-electron charging effects in small tunnel junctions. \textit{Phys. Rev. Lett.} \textbf{59,} 109-112 (1987).
\bibitem {Likharev1987}	
    Likharev, K. Single-electron transistors: electrostatic analogs of the DC SQUIDS. \textit{IEEE Trans. Magnetics} \textbf{23,} 1142-1145 (1987).
\bibitem {Kanda2004}%
	Kanda, A., Baelus, B. J., Peeters, F. M., Kadowaki, K. \& Ootuka, Y. (2004). Experimental evidence for giant vortex states in a mesoscopic superconducting disk. \textit{Phys. Rev. Lett.} \textbf{93,} 257002 (2004).
\bibitem{Schweigert1998}%
  Schweigert, V. A., Peeters, F. M. \& Singha Deo, P. Vortex Phase Diagram for Mesoscopic Superconducting Disks. \textit{Phys. Rev. Lett.} \textbf{81,} 2783-2786 (1998).
\bibitem{Palacios1998}%
  Palacios, J. J. Vortex matter in superconducting mesoscopic disks: Structure, magnetization, and phase transitions. \textit{Phys. Rev. B} \textbf{58,} R5948-R5951(R) (1998).
\bibitem {Giazotto2006}%
	Giazotto, F., Heikkil\"a, T. T., Luukanen, A., Savin, A. M. \& Pekola, J. P. Opportunities for mesoscopics in thermometry and refrigeration: Physics and applications. \textit{Rev. Mod. Phys.} \textbf{78,} 217 (2006).
\bibitem {Maisi2013}	
	Maisi, V. F. \textit{el al.} Excitation of single quasiparticles in a small superconducting Al island connected to normal-metal leads by tunnel junctions. \textit{Phys. Rev. Lett.} \textbf{111,} 147001 (2013).
\bibitem {Kopnin2001}%
	Kopnin, N. B. \textit{Theory of Nonequilibrium Superconductivity}. (Oxford Univ. Press, Oxford, 2001).
\bibitem{Aref2011}%
    Aref, T. \textit{el al.} Andreev tunneling in charge pumping with SINIS turnstiles.
  \textit{Europhys. Lett.} \textbf{96,} 37008 (2011).
\bibitem{Averin2008}
    Averin, D. V. \& Pekola, J. P. Nonadiabatic Charge Pumping in a Hybrid Single-Electron Transistor. \textit{Phys. Rev. Lett.} \textbf{101,} 066801 (2008).
\end{thebibliography}

%

\end{document}